\documentclass[twocolumn]{aastex63}

\usepackage{amsmath}

\shorttitle{Accretion-state transitions in Tau 4}
\shortauthors{Littlefield et al.}

\newcommand{\lsqfull}{LSQ~172554.8-643839}
\newcommand{\lsq}{LSQ~1725-64}
\newcommand{\mdot}{$\dot{M}$}

\begin{document}
	
\title{Short-cadence \textit{K2} observations of an accretion-state transition in the polar Tau 4}

\author[0000-0001-7746-5795]{Colin Littlefield}
\affiliation{University of Notre Dame, Notre Dame, IN 46556, USA}
\affiliation{Department of Astronomy, University of Washington, Seattle, WA 98195, USA}
\author[0000-0003-4069-2817]{Peter Garnavich}
\affiliation{University of Notre Dame, Notre Dame, IN 46556, USA}
\author[0000-0003-4373-7777]{Paula Szkody}
\affiliation{Department of Astronomy, University of Washington, Seattle, WA 98195, USA}
\author{Gavin Ramsay}
\affiliation{Armagh Observatory, College Hill, Armagh BT61 9DG, UK}
\author{Steve Howell}
\affiliation{NASA Ames Research Center, Moffett Field, CA 94035, USA}
\author[0000-0001-6013-1772]{Isabel Lima}
\affiliation{Department of Astronomy, University of Washington, Seattle, WA 98195, USA}
\affiliation{Instituto Nacional de Pesquisas Espaciais, S\~ao Jos\'e dos Campos, S\~ao Paulo, Brazil}
\author[0000-0001-6894-6044]{Mark Kennedy}
\affiliation{Jodrell Bank Centre for Astrophysics, School of Physics and Astronomy, The University of Manchester, Manchester M13 9P, UK}
\author{Lewis Cook}
\affiliation{CBA Concord, Concord, CA USA}
\affiliation{AAVSO observer}

\begin{abstract}
	
The \textit{Kepler} spacecraft observed a total of only four AM Herculis cataclysmic variable stars during its lifetime. We analyze the short-cadence \textit{K2} light curve of one of those systems, Tau 4 (RX~J0502.8+1624), which underwent a serendipitous jump from a low-accretion state into a high state during the final days of the observation. Apart from one brief flare, there was no evidence of accretion during the 70~d of observations of the low state. As Tau~4 transitioned into a high state, the resumption of accretion was very gradual, taking approximately six days ($\sim$90 binary orbits). We supplement Tau~4's \textit{K2} light curve with time-resolved spectroscopy obtained in both high and low states of accretion. High-excitation lines, such as He~II 468.6~nm, were extraordinarily weak, even when the system was actively accreting. This strongly suggests the absence of an accretion shock, placing Tau~4 in the bombardment regime predicted for AM Herculis systems with low accretion rates. In both the high-state and low-state spectra, Zeeman absorption features from the white dwarf's photosphere are present and reveal a surface-averaged field strength of $15\pm2$~MG. Remarkably, the high-state spectra also show Zeeman-split emission lines produced in a region with a field strength of $12\pm1$~MG. Zeeman emission has not been previously reported in an AM Herculis system, and we propose that the phenomenon is caused by a temperature inversion in the WD's atmosphere near the accretion region.
	
\end{abstract}


\section{Introduction}

    \subsection{Polars}

	The \textit{Kepler} satellite has provided seminal insights into a wide range of cataclysmic variable stars (CVs), but the resulting literature has generally focused on CVs that harbor non- or weakly magnetic white dwarfs (WDs). CVs containing strongly magnetized ($B\gtrsim 10$~MG) WDs have received comparatively little attention. These objects are known equivalently as polars and AM Herculis systems, and their high field strengths cause them to have fundamentally different properties than other CVs \citep[for a review, see][]{cropper}. Perhaps most famously, the WD's magnetic field forces the WD's rotational period to synchronize with the binary orbital period. Another critical property of polars is the absence of an accretion disk; the accretion flow couples to the WD's magnetic-field lines shortly after leaving the companion star, so accretion occurs on a free-fall timescale rather than the much longer viscous timescale of an accretion disk. Consequently, variations in a polar's mass-transfer rate have an immediate effect on the system's accretion luminosity. Long, uninterrupted light curves of polars therefore offer an excellent opportunity to assess the stability of the mass-transfer rate.

	Before material can accrete onto the WD, it must usually pass through a hydrodynamic shock, and the post-shock material cools by emitting a combination of optical/near-infrared cyclotron radiation and X-ray brehmstrahhlung. Cyclotron radiation is strongly beamed, showing maximum intensity when the observer's line of sight is perpendicular to the emission region, and it can be the dominant source of optical luminosity in accreting polars. For systems with low mass-transfer rates, cyclotron emission can cool infalling matter on the scale length of the ions' mean free path without a shock forming, a scenario known as the ``bombardment solution'' \citep{kp82}.  While the particle collisions do heat the WD photosphere, the temperature is expected to be substantially less than the temperature of shock-heated gas. Systems in the bombardment regime should therefore emit significantly less ionizing radiation than a polar with a shock.
	
	The mass-transfer rate (\mdot) varies enormously in many polars, causing these systems to meander unpredictably from states of vigorous \mdot\ to prolonged low states during which \mdot\ might plummet to negligible levels. These low states are often presumed to be related to changes on the secondary. The formation of starspots is one widely invoked model, since a starspot at the inner Langrangian point would suppress the scale height of the local photosphere and thereby reduce Roche-lobe overflow \citep{lp94, kc98}. Another proposal is that in short-period systems, the secondary's photosphere significantly underfills the star's Roche lobe, and the secondary's chromospheric activity determines whether Roche-lobe overflow occurs \citep{howell00}. Observational tests of these theories are difficult because accretion-state transitions occur unpredictably. 
	
	The continuous, extended photometry provided by the \textit{Kepler} spacecraft offers a new resource for understanding polars. \textit{Kepler} observed four known polars, all during its \textit{K2} mission. Although many more polars have been (or will be) visible to the Transiting Exoplanet Survey Satellite (TESS) due to its much greater sky coverage, \textit{Kepler} had the advantage of a significantly larger aperture, longer stare times, and a better plate scale, making it more suitable for the study of low-state polars.
		
	Of the four \textit{Kepler} polars, only one, EU Cnc \citep{hill}, has been previously reported. EU Cnc was observed in two campaigns, and \citet{hill} showed that its orbital profile remained extraordinarily stable during both campaigns, suggesting that there was no significant variation in the accretion rate during those observations. The authors argued that the WD likely was likely accreting from the wind of the secondary and that the system might be a low-accretion-rate polar \citep{schmidt}. Additionally, one TESS observation of a polar (CD Ind) has been published to date \citep{hakala, littlefield}.
	
	Two other polars, SDSS~J092122.84+203857.1 and V358~Aqr, round out the census of AM Herculis systems observed by \textit{Kepler}. We will report the SDSS~J092122.84+203857.1 observations in a forthcoming research note, and a separate group is analyzing the V358~Aqr data.

	\subsection{Tau 4}

	\begin{figure}
		\centering
		\includegraphics[width=\columnwidth]{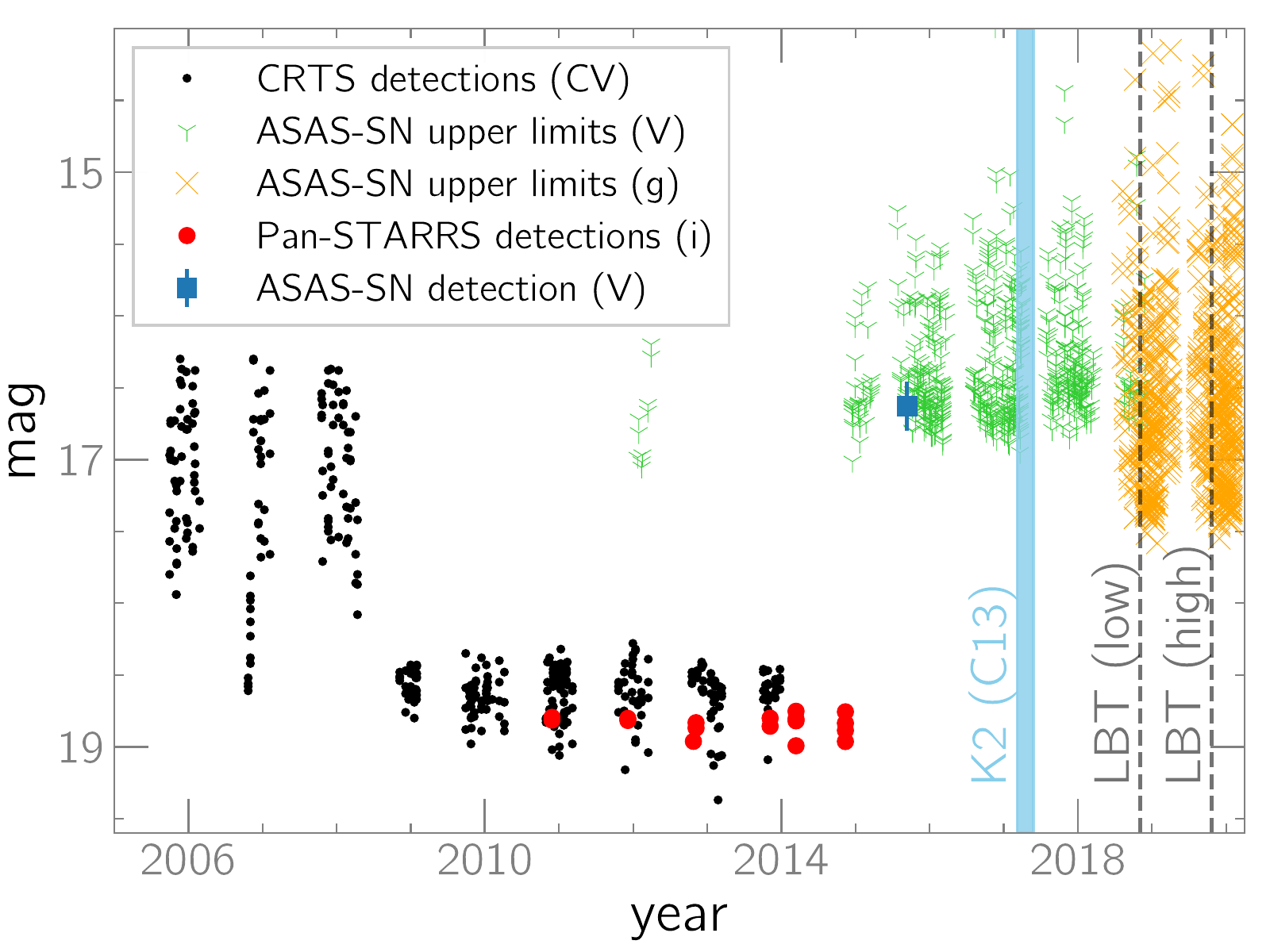}
		\caption{The long-term light curves of Tau 4, assembled from the CRTS, ASAS-SN, and Pan-STARRS. Note the different bandpasses in the legend. The shaded blue region and the dashed vertical lines indicate, respectively, the epochs of observation for the \textit{K2} light curve in 2017 and the LBT spectra in 2018 and 2019.  The CRTS data show that Tau 4 was in a state of active accretion from 2005-2008 before settling into a prolonged low state. However, the non-detection by ASAS-SN of the high state during the 2019 LBT spectra establishes that Tau~4's high states can fall below the ASAS-SN detection threshold. The one apparent ASAS-SN detection is likely spurious because of its modest SNR (6.5) and because Tau~4 was undetected in exposures obtained just 109~s and 218~s earlier. \label{fig:longterm} }
	\end{figure}

	Because of its X-ray emission and its optical spectrum, \citet{motch} flagged Tau 4 (= RX J0502.8+1624) as a likely magnetic CV. Its near-IR colors indicated a possible brown-dwarf donor star, a possibility investigated by \citet{littlefair} with a single near-IR spectrum. Instead of detecting a brown dwarf, they found that Tau 4's K-band spectrum was dominated by cyclotron radiation and that this was the cause of the system's unusual near-IR colors. 
	
	\citet{howell} followed up on this result by obtaining phase-resolved IR spectroscopy and optical photometry of the system. Their photometry revealed large-amplitude ($>$ 2 mag) variability on timescales of $\sim$20 min, but they did not find an obvious periodicity. Spectroscopically, they detected significant cyclotron radiation as well as a brief inversion of the spectral lines, which switched from emission to absorption and back in just $\sim$6~minutes. They attributed the line inversion to a viewing geometry in which line absorption by cool gas in the accretion flow obscures the cyclotron-emitting region at a particular phase in each orbital cycle, an effect that has been reported in at least five other polars \citep[see Sec. 4.2.1 in ][and references therein]{j1321}. As pointed out in \citet{j1321}, three of those five other polars are eclipsing systems, so the presence of the line-inversion phenomenon might imply a relatively high orbital inclination.
	
	More recently, \citet{harrison} analyzed a new set of near-IR spectra, and by modeling the cyclotron harmonics, he measured a magnetic field strength of $B= 12.3$~MG and a shock temperature of $kT = 14$~keV. He also identified two possible orbital periods: 1.55~h or 3.11~h, the latter of which was slightly preferred.

	Publicly available survey photometry of Tau~4 by the Catalina Real-Time Transient Survey \citep[CRTS;][]{drake}, the All-Sky Automated Survey for Supernovae \citep[ASAS-SN;][]{shappee, kochanek}, and Pan-STARRS1 \citep{chambers} shows that Tau~4 was in a high-accretion state between 2005 and early 2008 (Fig.~\ref{fig:longterm}). Although the survey photometry has not detected a high state in Tau~4 since 2008, Fig.~\ref{fig:longterm} shows that this can be attributed, at least in part, to insufficient depth and sampling of recent surveys.

	Tau 4 has a Gaia DR2 parallax of $\pi = 4.75\pm0.36$~mas \citep{gaia, dr2}, corresponding to a distance of $211^{+18}_{-16}$ pc \citep{bj18}.

\section{Data}

	\subsection{Kepler} \label{sec:kepler}

    \begin{figure}
        \centering
        \includegraphics[width=\columnwidth]{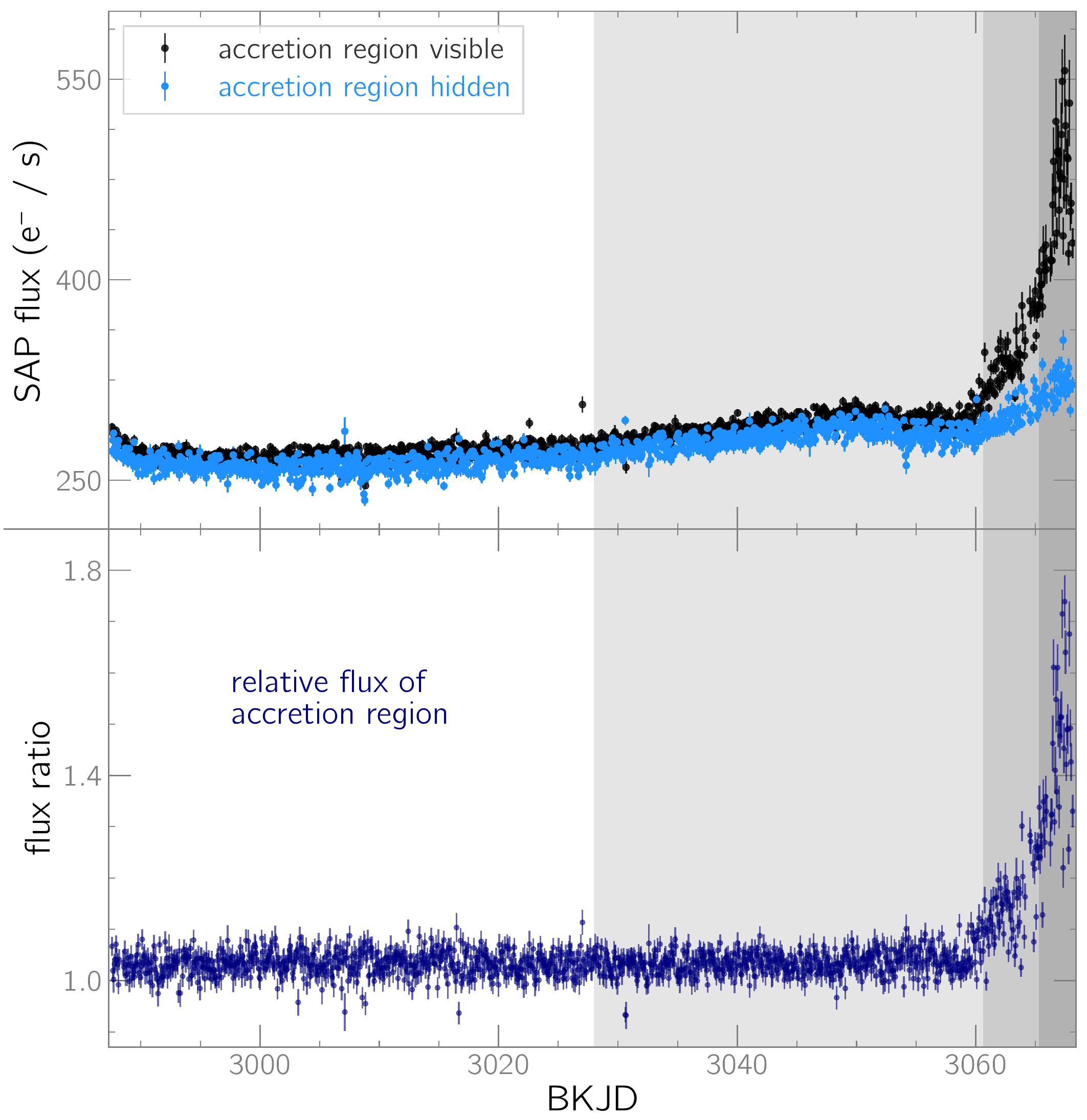}
        \caption{Light curves showing the accretion state of Tau~4 during the \textit{K2} observation. In the top panel, we show, for each individual binary orbit, the average brightness during the part of the orbit when the accretion region is visible (black points) and the part of the orbit when it is not (blue points). The bottom panel plots the ratio of these two measurements, and the rise at the end indicates the resumption of accretion. The four horizontal shaded regions show the time bins used in Fig.~\ref{fig:orbital_waveforms}.}
        \label{fig:cyclotron_lc}
    \end{figure}

	\textit{Kepler} observed Tau 4 for 80.6 days between 2017 March 8 and 2017 May 27, executing the observations at its short cadence of 58.8 s. We used {\tt lightkurve} to extract the light curve from the target-pixel file (TPF), rejecting any observations with non-zero quality flags. We also created light curves of the TPF background and confirmed that there were no significant systematic trends.

	The extracted light curve was peppered with outliers, primarily from cosmic-ray hits. Since cosmic-ray hits inside the aperture mask can cause a sudden jump in the centroid of a moderately faint source like Tau~4, we measured Tau~4's centroid in each frame and excluded frames that showed large, abrupt jumps. We then inspected the remaining brightening events in the light curves to determine whether they could be attributed to obvious artifacts in the TPFs, such as spurious brightenings of the background. The centroid information further enabled us to correlate Tau~4's flux with its position on the sensor. We identified and mitigated a gradual rise in flux that correlated with Tau~4's centroid.
	
	When describing the \textit{K2} light curve, we adhere to the BKJD convention, defined as BKJD = BJD - 2454833, to describe event times.

	\subsection{LBT Spectroscopy}
	
	The Large Binocular Telescope (LBT) observed Tau~4 with the MODS spectrographs \citep{pogge} on 2018 November 5 and 2019 October 23. The 2018 data were obtained in terrible conditions, with highly variable cloud cover, but conditions during the 2019 observations were ideal. The MODS spectrographs provide wavelength coverage from 300-1000 nm, with a dichroic at 575~nm. Both sets of spectra utilized a 0.8-arcsec slit, the G670L 250 lines mm$^{-1}$ grating for the red spectra, and the G400L grating 400 lines mm$^{-1}$ for the blue spectra. Exposures were 200~s. The spectra were reduced using standard procedures in IRAF, but the cloud cover during the 2018 data makes the flux calibration unreliable.

	\subsection{Ground-based time-series photometry}
	
	In support of the 2019 LBT spectroscopy, we obtained time-series photometry with three ground-based telescopes: the 80-cm Sarah L. Krizmanich Telescope on the campus of the University of Notre Dame and a pair of 51-cm Planewave telescopes, one located at the Sierra Remote Observatory and the other at the New Mexico Skies remote observatory.	Observations took place on 2019 November 4, 6, 16, 17, 19, and 22 and were unfiltered with a Johnson $V$ zeropoint.

	\begin{figure}
		\centering
		\includegraphics[width=\columnwidth]{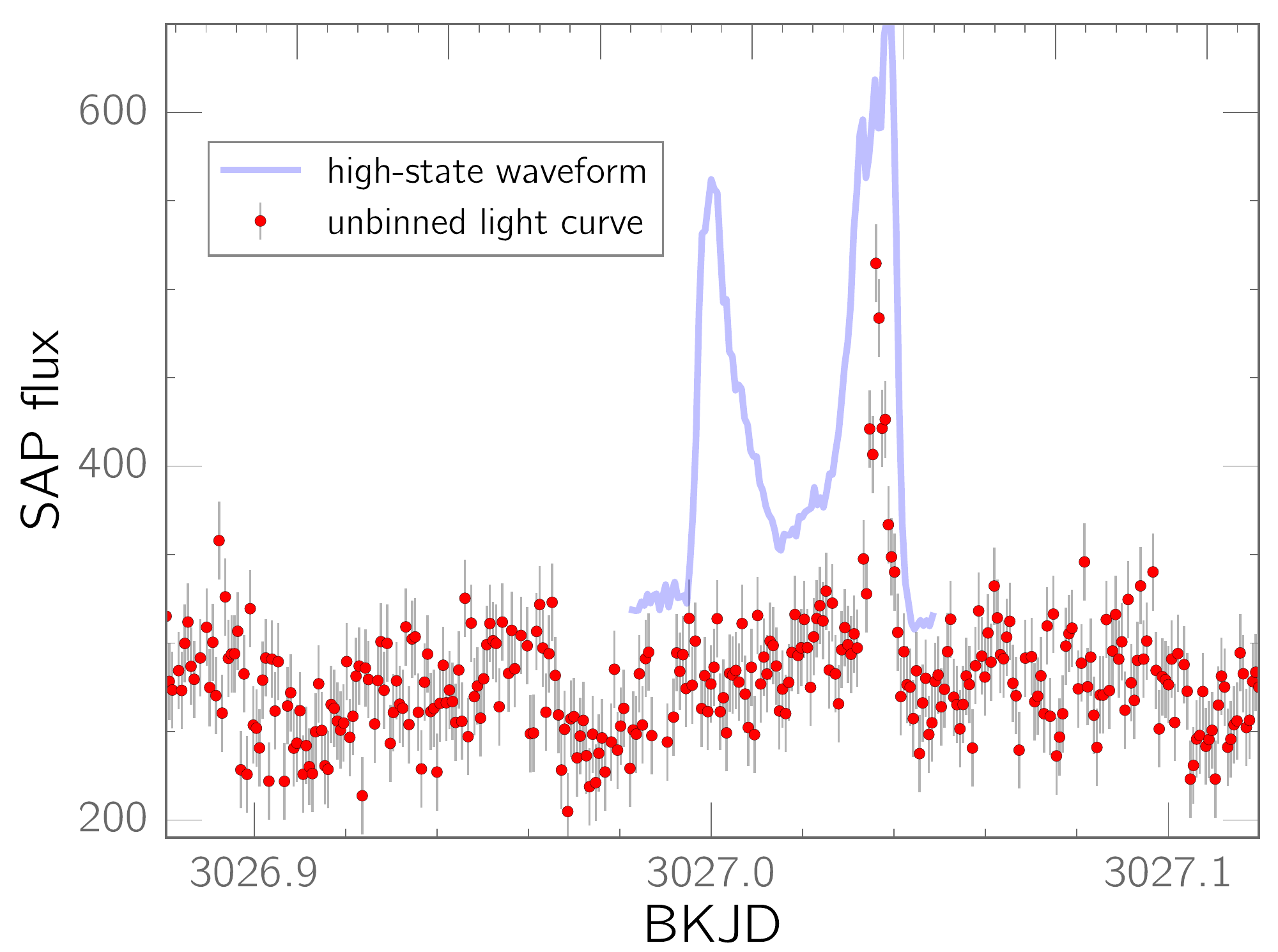}
		\caption{Light curve of a flare observed during the low state. One orbital cycle of the phase-averaged high-state profile is plotted for reference in order to establish that the flare occurred at an orbital phase during which the accretion region is visible. Inspection of the target-pixel file confirmed that the flare is not attributable to instrumental artifacts or the passage of a minor planet through the photometric aperture.
		\label{fig:flare}}
	\end{figure}

	\begin{figure*}
		\centering
		\includegraphics[width=\textwidth]{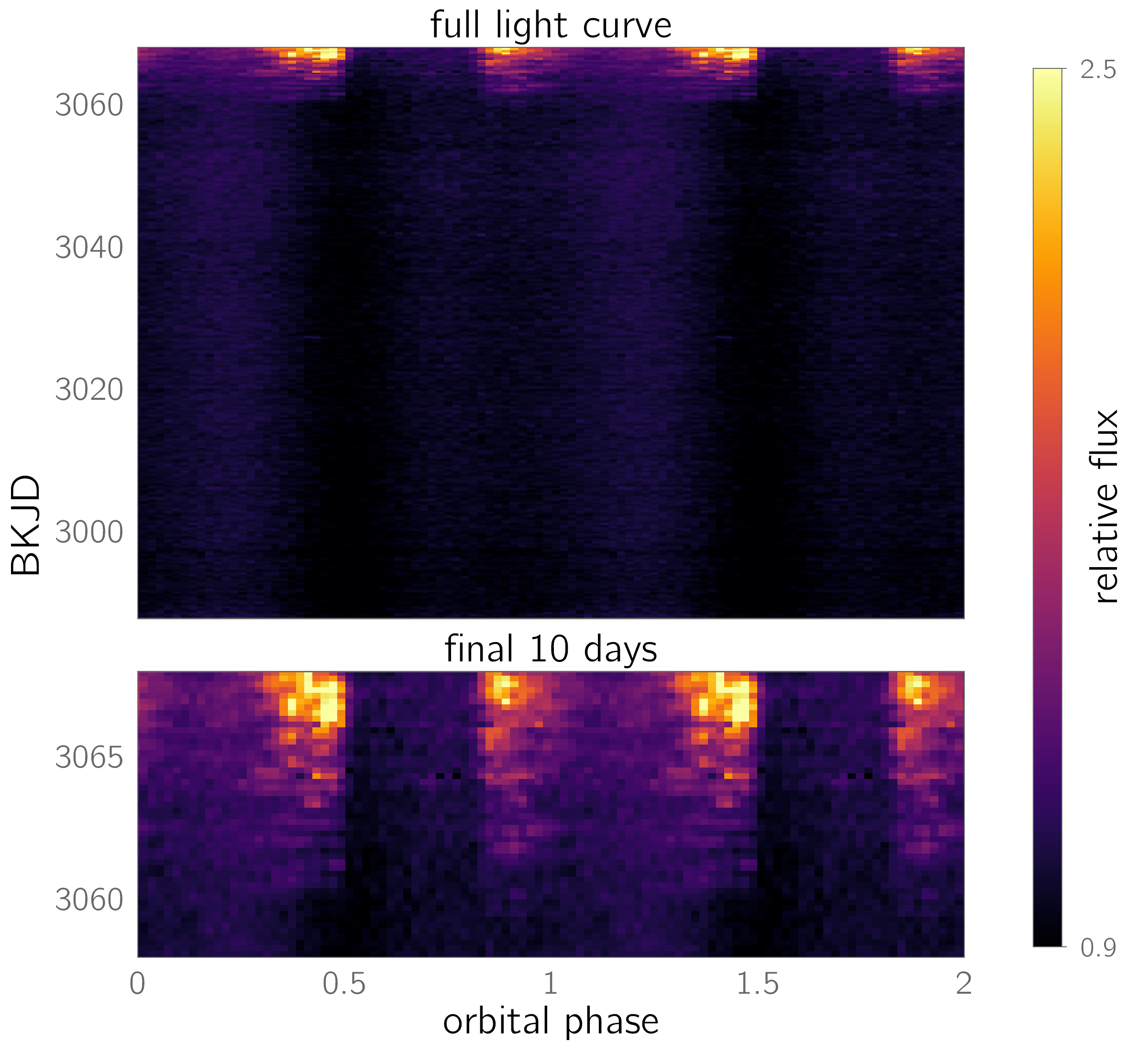}
		\caption{Two-dimensional light curve, phased to the orbital period. The top panel shows the entire light curve, while the bottom enlarges the final 10 days, when accretion resumed. Both panels use a 0.4-day sliding window with 50 non-overlapping bins per orbit.
		\label{fig:lightcurve2D} }
	\end{figure*}

    \begin{figure}
		\includegraphics[width=\columnwidth]{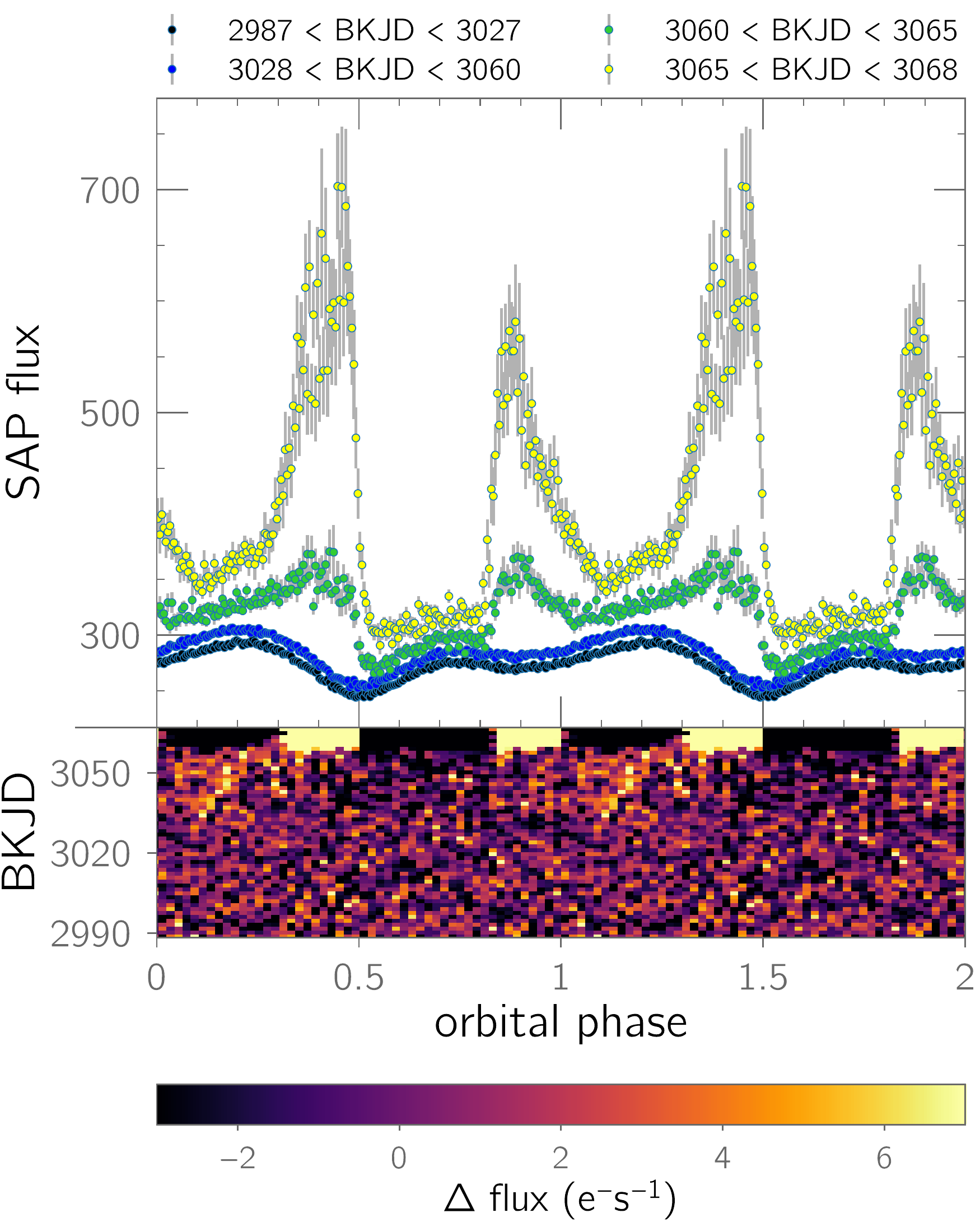}
		\caption{ {\bf Top:} The phase-averaged orbital profile of Tau 4 during the four segments of the \textit{K2} light curve indicated by the shaded regions in Fig.~\ref{fig:cyclotron_lc}. {\bf Bottom:} Folded, two-dimensional light curve after subtracting the phase-averaged profile of the earliest bin. The mean of each row was subtracted to increase the visibility of low-contrast features. In the $\sim30$~d prior to the start of detectable accretion, there was a subtle brightening during the orbital phases corresponding with the best visibility of the then-dormant accretion region. This might suggest an increased hotspot temperature on the WD, possibly a result of the flare at BKJD=3027 from Fig.~\ref{fig:flare}.
		\label{fig:orbital_waveforms}}
	\end{figure}

\section{The \textit{K2} light curve}

	During the first 70 days of Campaign 13, Tau~4 was mired in a low-accretion state, but it brightened dramatically during the final 10 days of the observation (Fig.~\ref{fig:cyclotron_lc}). The low state was characterized by a stable, double-humped profile that we attribute to the binary orbital period, and with the exception of a single flare observed near BKJD = 3027 (Fig.~\ref{fig:flare}), there were no unambiguous accretion events during the low state.

	Using the period of the low-state variation, we compute an orbital ephemeris of
	\begin{equation} 	\label{ephemeris}
	T_{conj} = 2457860.53915(63) + 0.06636579(23)\times E
	\end{equation}
	where $T_{conj}$ is the predicted time of inferior conjunction expressed in Barycentric Julian Date in Barycentric Dynamical Time. The $T_0$ term was determined from radial-velocity measurements of the secondary's near-infrared Na~I absorption doublet (see Sec.~\ref{sec:spectra2019}). The numbers in parentheses are $1\sigma$ uncertainties on the final two digits of the corresponding parameters. The orbital period of $\sim$1.59~h is consistent with the shorter of the two candidate periods (1.55~h) from \citet{harrison}.

	We use Eq.~\ref{ephemeris} to construct a phased, two-dimensional light curve that showcases a dramatic change in Tau~4's orbital profile during the final 10 days of the \textit{K2} light curve. Fig.~\ref{fig:lightcurve2D} shows that during that time, the profile began to show erratic accretion events whose frequency and amplitude gradually increased. Equally striking in Fig.~\ref{fig:lightcurve2D} is the absence of accretion events during the same 30\% of each orbit throughout the high state. This feature, commonly referred to as a ``self-eclipse'' in other polars, occurs when the accretion region rotates behind the limb of the WD.
	
	Fig.~\ref{fig:orbital_waveforms} provides a more traditional visualization by plotting the phase-averaged orbital profile from four separate bins, corresponding with the four differently-shaded regions of the binned light curve in Fig.~\ref{fig:cyclotron_lc}. When accretion resumed, the smooth, double-humped orbital modulation quickly gave way to a jagged, M-shaped profile. This a classic cyclotron-beaming effect; cyclotron radiation is brightest when the observer's viewing angle is perpendicular to the accreting field line, and its intensity decreases as the viewing angle becomes increasingly parallel to the field line \citep[e.g.][]{gansicke}. For Tau~4, maximum cyclotron beaming occurs almost immediately when the cylcotron-emitting region rotates in into view, and the observed cyclotron flux diminishes as the accretion region approaches the meridian of the WD. 

    The earliest two orbital profiles in Fig.~\ref{fig:orbital_waveforms} cover two segments of the low state, with the second bin encompassing the gradual rise seen in Fig.~\ref{fig:cyclotron_lc}. A comparison of these two profiles (bottom panel of Fig.~\ref{fig:orbital_waveforms}) establishes that there was an extremely subtle change in the orbital profile during the $\sim$30~d preceding the resumption of obvious accretion, with a low-amplitude brightening centered on orbital phase $\sim$0.15, which is also the approximate midpoint of the cyclotron-bright part of the orbit during the high state. This suggests that the low-state brightening might be caused by heating of the accretion region---perhaps as a result of the flare in Fig.~\ref{fig:flare}. The low-state brightening is unlikely to be from cyclotron radiation, as there is no sign of the cyclotron beaming effects observed during the high state.

	\begin{figure*}
		\centering
		\includegraphics[width=\textwidth]{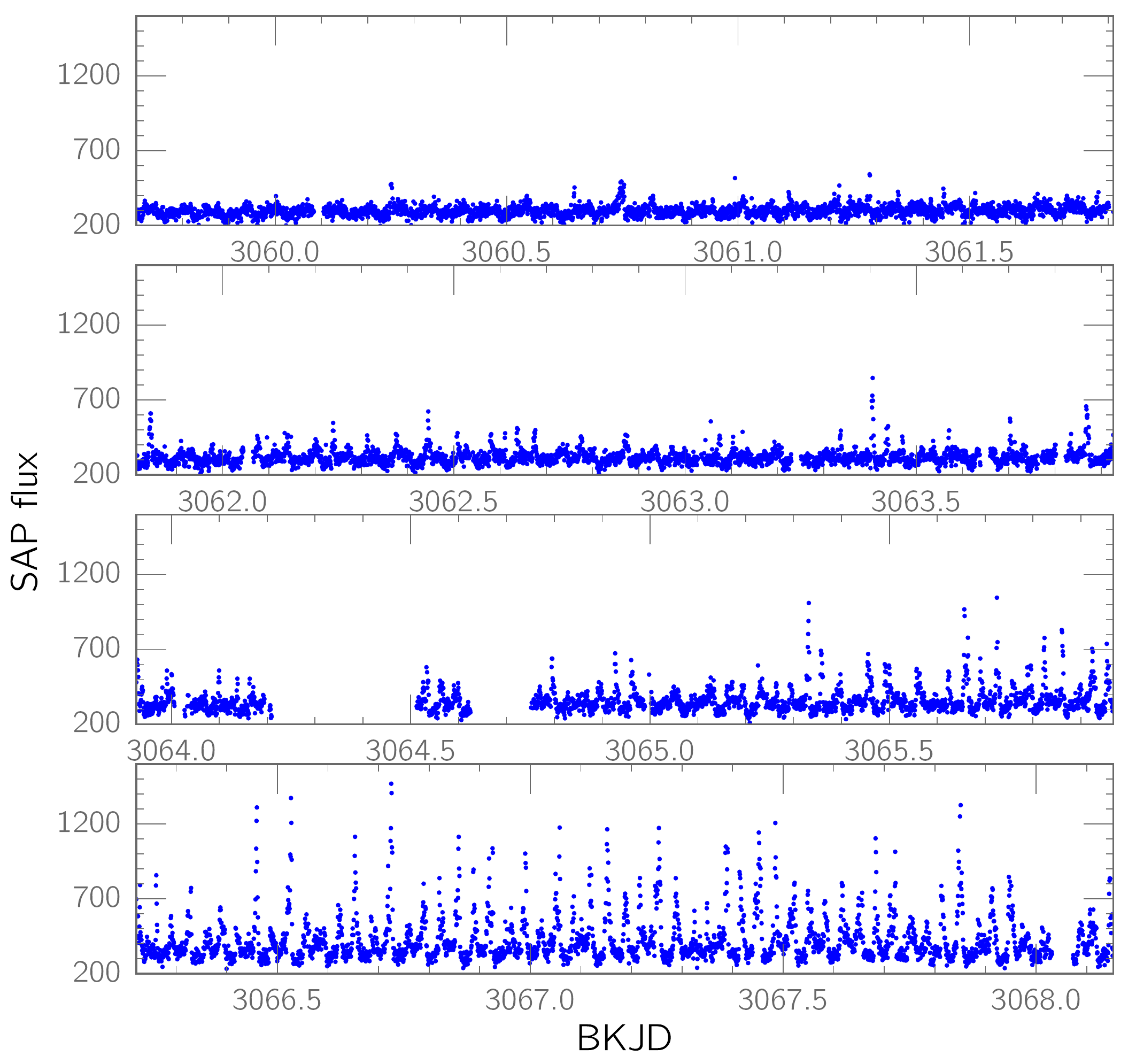}
		\caption{The unbinned light curve during the transition from the low state into the high state. The integration time for each point is one minute. Each panel uses the same horizontal and vertical scaling in order to enable direct comparisons between them. Sporadic accretion resumed after BKJD = 3060 and became frequent after BKJD = 3062. Around BKJD = 3065.5, the amplitude of the accretion flares increased significantly and remained high until the end of the campaign. \label{fig:transition}}
	\end{figure*}

	\subsection{The transition into a high state}
	
	The \textit{K2} light curve offers an unprecedented opportunity to study the resumption of accretion after a low state with negligible accretion. Fig.~\ref{fig:transition} presents the unbinned light curve during this transition. The first unambiguous sign of accretion occurred with a brief brightening at BKJD = 3060.75. There is some evidence of a pair of three-minute flares in the preceding 12 hours. We could not identify any obvious artifacts that would produce them, and they were observed at orbital phases during which the cyclotron-emitting region is visible. The transition into the high state therefore began somewhere between BKJD = 3060.2-3060.75.
	
	Over the course of the next six days, erratic brightening events became increasingly common, and their amplitude also tended to increase (Figs.~\ref{fig:lightcurve2D}~and~\ref{fig:transition}). Because these events were observed exclusively when the accretion region was visible, they were almost certainly bursts of cyclotron radiation. In contrast, stellar flares by the secondary would not show this orbital-phase dependence.

	The unpredictable nature of the light curve makes it difficult to objectively identify when Tau~4 entered its high state, but the amplitude of the cyclotron pulses began to level off at BKJD $\sim$ 3066.4. The full transition therefore occurred during a six-day span ($\sim$90 binary orbits).

\section{LBT spectroscopy}

	\begin{figure}
		\centering
		\includegraphics[width=\columnwidth]{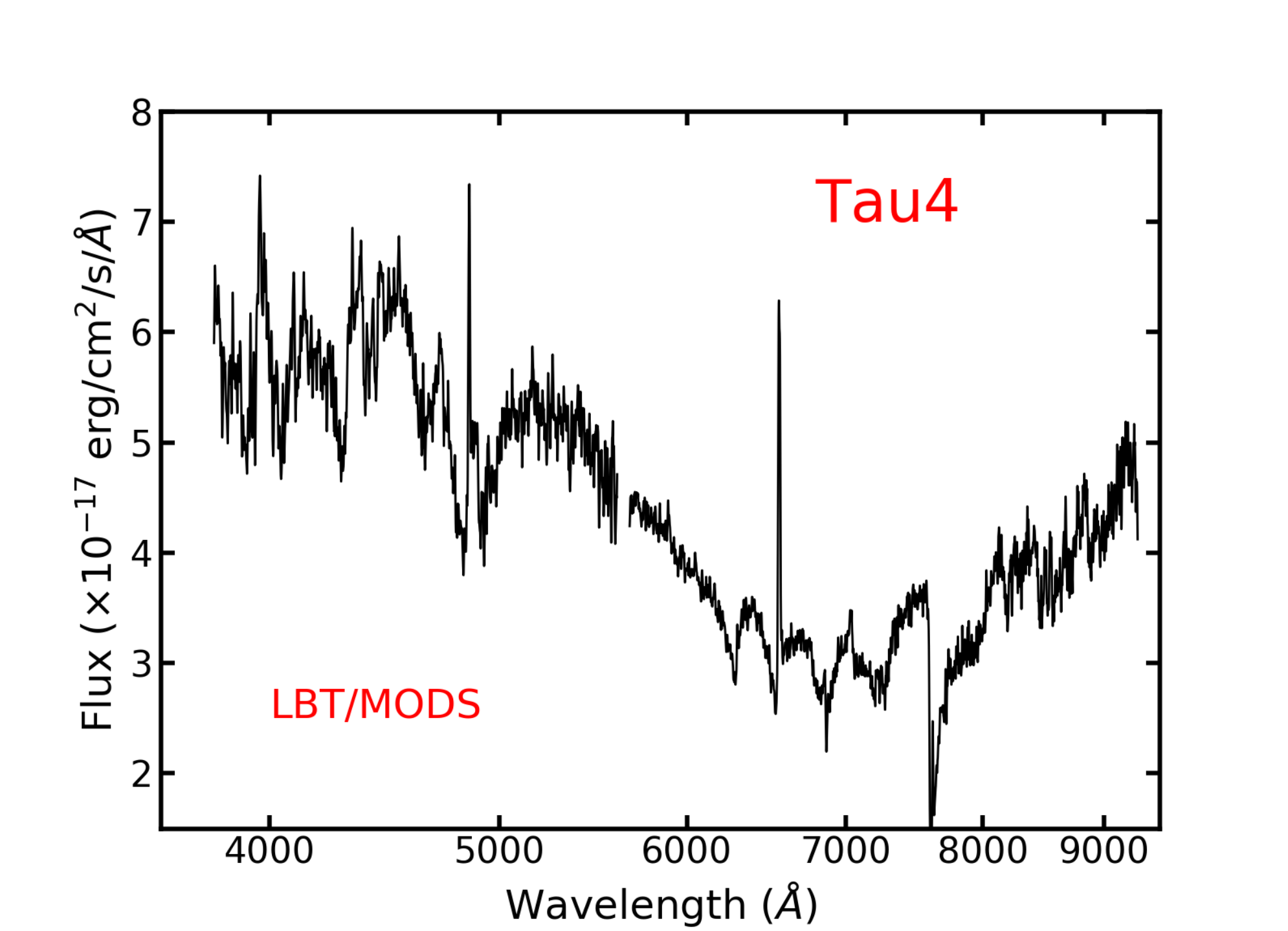}
		\caption{LBT spectrum of Tau 4 in a low state in 2018. The WD dominates at wavelengths shorter than $\sim$7000\AA\ and displays Zeeman splitting at the Balmer lines. The two main emission lines are H$\alpha$ and H$\beta$. The Ca~II near-infrared triplet is weakly present in emission. \label{fig:spectrum} }
	\end{figure}

    \begin{figure}
        \centering
        \epsscale{1.15}
        \plottwo{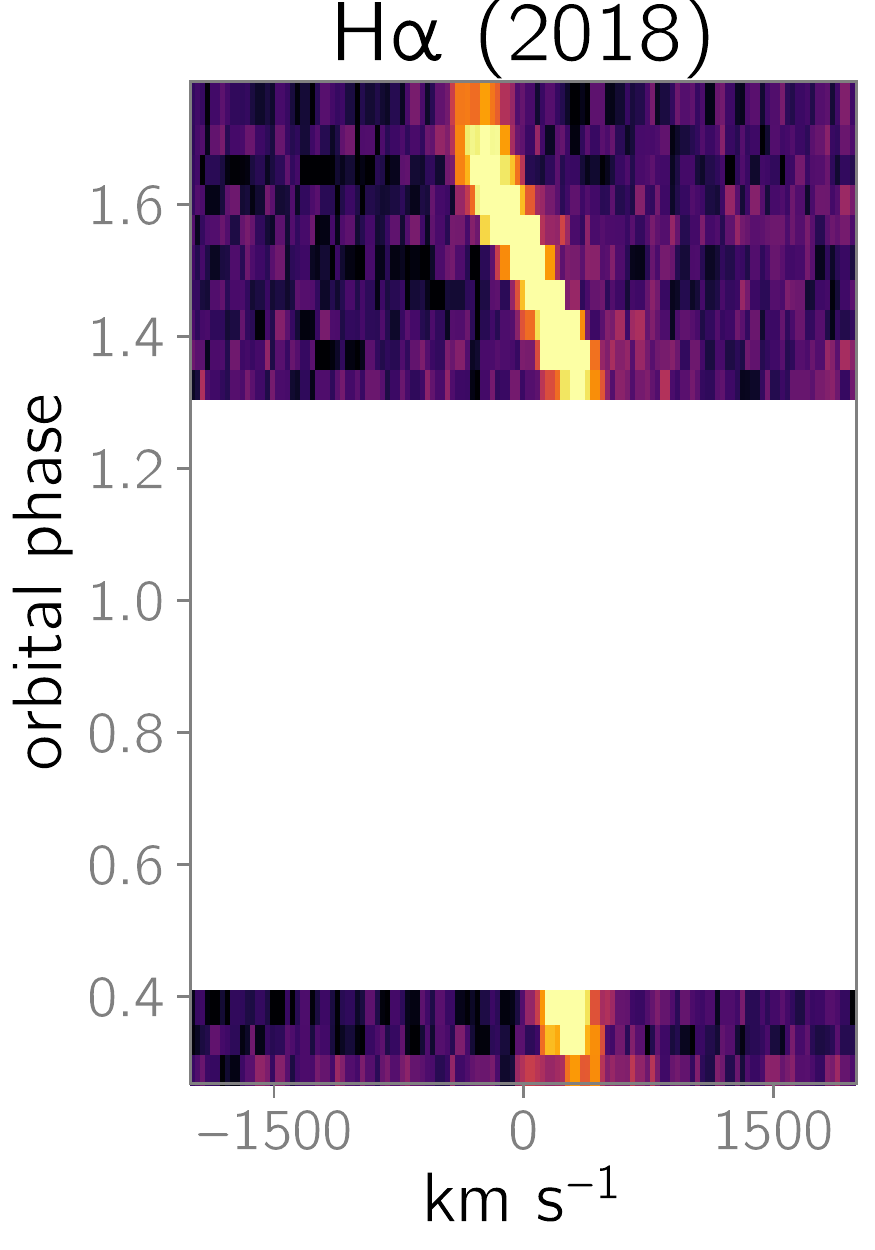}{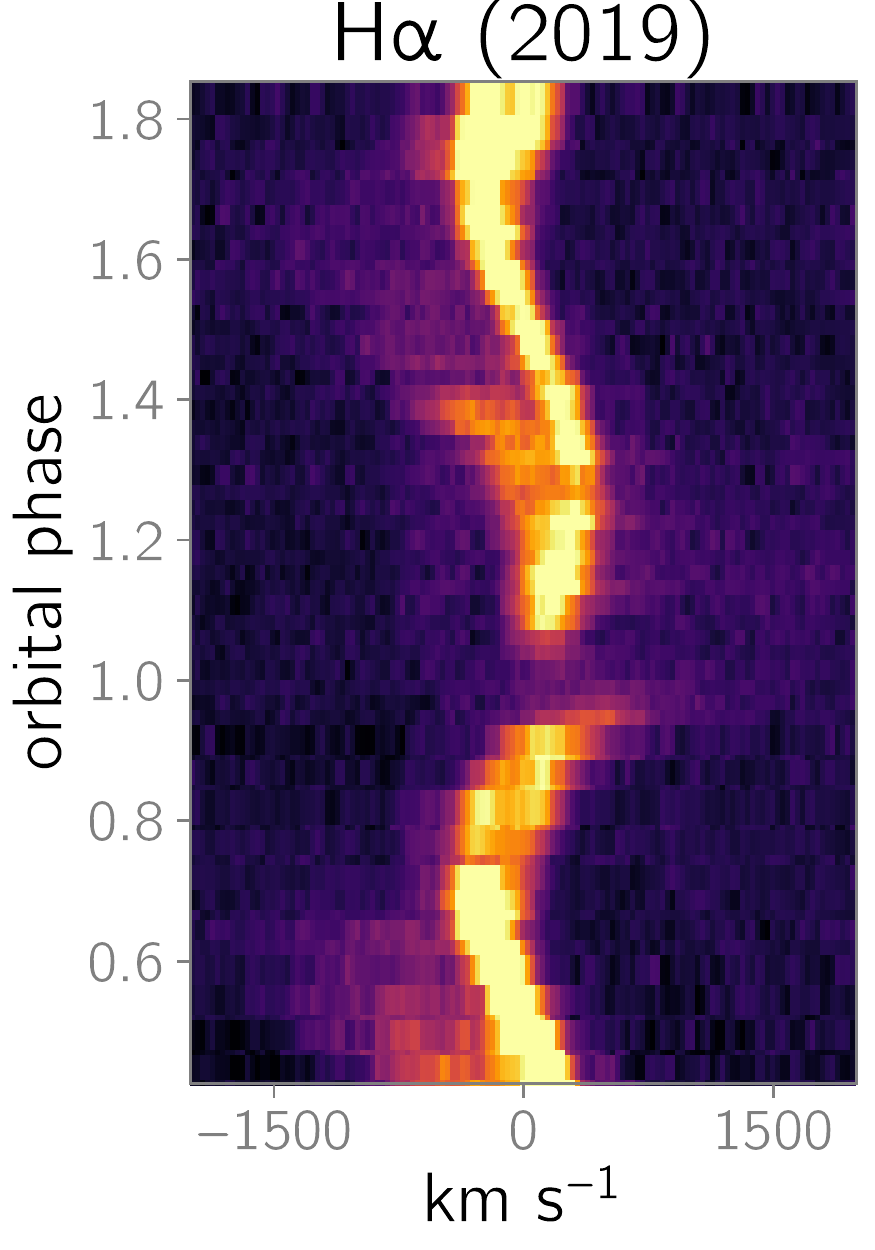}
        \caption{Trailed, continuum-subtracted spectra of the H$\alpha$ line in the 2018 and 2019 LBT spectra. The gap in the 2018 dataset is due to cloud cover. Low-velocity emission from the donor is visible in both datasets, while there is an additional high-velocity component in the high-state spectra from 2019. The low-velocity emission disappears at inferior conjunction.}
        \label{fig:trailed_halpha}
    \end{figure}

    \subsection{2018 low state}

	The 2018 LBT spectrum of Tau 4 (Fig.~\ref{fig:spectrum}) consists primarily of the photospheres of WD and the companion star, along with strong emission at the H$\alpha$ and H$\beta$ lines and weak emission at the Ca~II near-infrared triplet. Fig.~\ref{fig:trailed_halpha} shows a trailed spectrum of H$\alpha$ and establishes that the line consisted solely of a narrow component from the inner hemisphere of the secondary, with a projected velocity semi-amplitude of 300 km s$^{-1}$. The absence of He emission lines and of high-velocity components to the Balmer lines indicate that the accretion rate was negligible during the observations, similar to the \textit{K2} low state. The WD's spectrum shows obvious Zeeman splitting of the Balmer absorption lines. Unfortunately, poor observing conditions caused nearly a full orbit's worth of spectra to be unusable, precluding a more in-depth analysis. 
	
	\subsection{2019 high state}
	\label{sec:spectra2019}

    \begin{figure}
        \centering
        \includegraphics[width=\columnwidth]{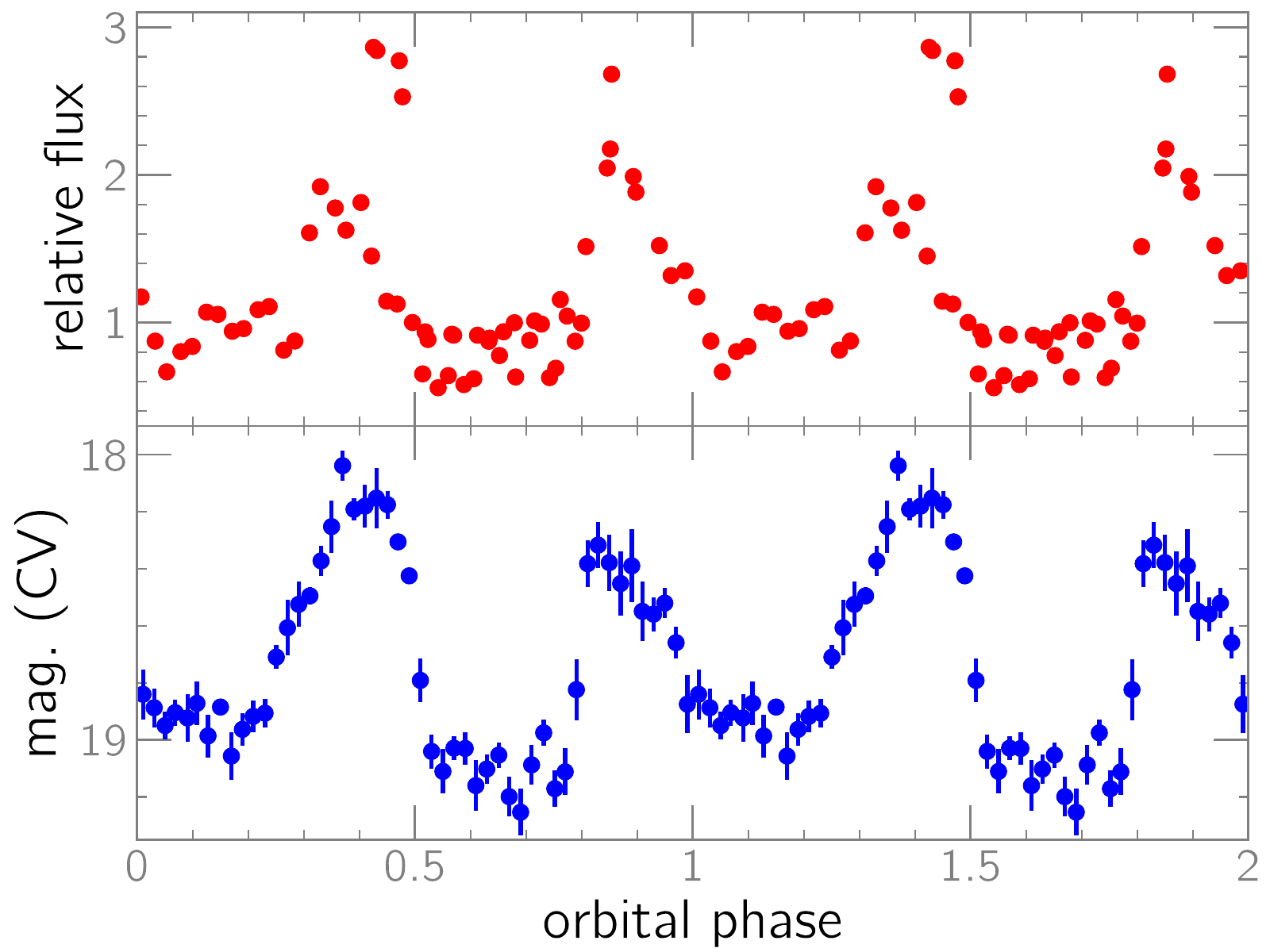}
        \caption{Phased light curves of Tau~4 near the time of the 2019 LBT spectra. {\bf Top:} Pseudo light curve during the LBT observations, created by integrating the flux of the red spectra. {\bf Bottom: } Light curve from ground-based photometry obtained on 2019 November 6, fourteen days after the LBT observation. The `CV' bandpass is unfiltered with a Johnson $V$ zeropoint. The profiles are identical to the one observed during the \textit{K2} high state, suggesting a similar accretion state in both datasets. An additional light curve from 2019 November 4 also showed Tau~4 to be in a high state.  
        }
        \label{fig:lightcurve2019}
    \end{figure}
    
    Fortunately, the observing conditions during the 2019 LBT spectra, when Tau~4 was in a high state, were much more favorable. The integrated flux of the spectra and ground-based light curves obtained 12 and 14 days after the LBT spectra establish that the orbital profile was identical to that observed during the \textit{K2} high state (Fig.~\ref{fig:lightcurve2019}). This, in turn, suggests a nearly identical accretion state in the \textit{K2} and 2019 LBT observations.
    
    Most fundamentally of all, the 2019 spectra enable us to determine the absolute orbital phase of Tau~4 at any point during the \textit{K2} light curve. The Na~I absorption doublet at $\lambda\lambda$8183, 8195\AA\ is readily apparent in the spectra and is unambiguously attributable to the donor star's photosphere. Unlike the secondary's narrow emission-line components, which disappear for part of the orbit, the Na~I absorption is observable at all orbital phases, making it a more reliable tracer of the secondary's motion. The radial-velocity curve of the Na~I doublet has a projected semi-amplitude of 440 km s$^{-1}$, and the phasing of this feature enables us to measure the time of inferior conjunction, providing the $T_0$ term in the orbital ephemeris (Eq.~\ref{ephemeris}). 
 
    Two of the most fundamental differences between the 2018 and 2019 spectra are the presence of He~I emission and a high-velocity ($\sim$2000 km s$^{-1}$) component in both the Balmer and He~I lines. The high-velocity component shows maximum redshift at inferior conjunction (Fig.~\ref{fig:trailed_halpha}), which suggests radial inflow toward the WD from the vicinity of the L1 point. We explore this in more detail with Doppler tomography in Sec.~\ref{sec:tomography}.

    The emission-line spectrum of Tau~4 is unusual for an accreting polar and differs from a 1995 high-state spectrum of the system \citep{motch}. Although strong He~I and He~II emission is a common characteristic of an accreting polar, these lines were extraordinarily weak in the LBT high-state spectra, with He~II $\lambda4686$\AA\ being absent altogether. The most straightforward explanation for this observation is that there was no accretion shock during the 2019 spectra, which is possible only if Tau~4 were accreting in the bombardment regime. Because of the detection of He~II emission during a high state in 1995 \citep{motch}, we conclude that Tau~4 has experienced at least two types of high states during its observational history: one in which there is no shock (the low-\mdot\ bombardment regime), and another in which there is (higher \mdot).
    
    Other notable emission features include the chromospheric Ca~II triplet and the Na~D doublet at $\lambda\lambda$5890, 5896\AA. The phase and amplitude of their radial-velocity variations agree with the narrow component seen in the Balmer and He~I lines. However, unlike the Balmer and He~I lines, there is no high-velocity component to the Ca~II and Na~D emission.
    
    All detectable emission lines contain a low-velocity (310 km s$^{-1}$), narrow component that (1) is in-phase with the Na~I absorption and (2) disappears at the secondary's inferior conjunction (Fig.~\ref{fig:trailed_halpha}). The narrow component must therefore originate on the inner hemisphere of the secondary. Furthermore, in order to produce such a strong orbital modulation in the line flux from this region, Tau~4 must have a moderately high orbital inclination.
    
\begin{figure}
        \centering
        \includegraphics[width=\columnwidth]{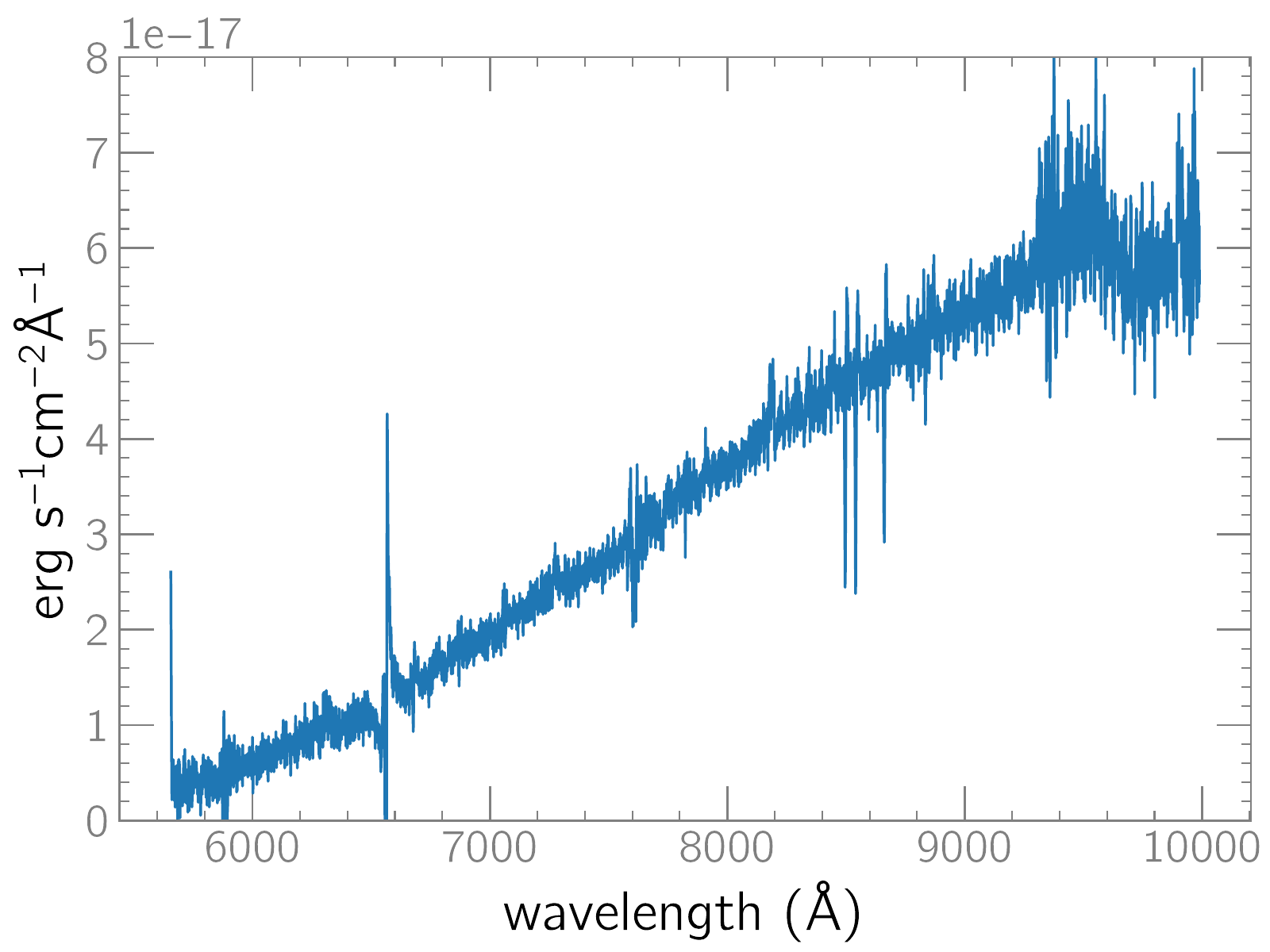}
        \caption{Average cyclotron spectrum of Tau~4, created by subtracting the cyclotron-free spectrum ($0.5 < \phi_{orb} < 0.8$) from the cyclotron-bright spectrum (the rest of the orbit). The spikes and dips near H$\alpha$ and the Ca~II triplet are artifacts from the subtraction process. The cyclotron spectrum is smooth and featureless, consistent with the higher-order harmonics blending together.}
        \label{fig:cyclotron_spec}
    \end{figure}

    In Fig.~\ref{fig:cyclotron_spec}, we isolate the cyclotron spectrum by subtracting the average spectrum during the cyclotron-faint part of the orbit from the average spectrum during the remainder of the orbit. No individual harmonics are visible, likely because the higher-order harmonics are smeared together. 
    
    \subsection{Doppler tomography} \label{sec:tomography}
    
\begin{figure}
        \centering
        \includegraphics[width=\columnwidth]{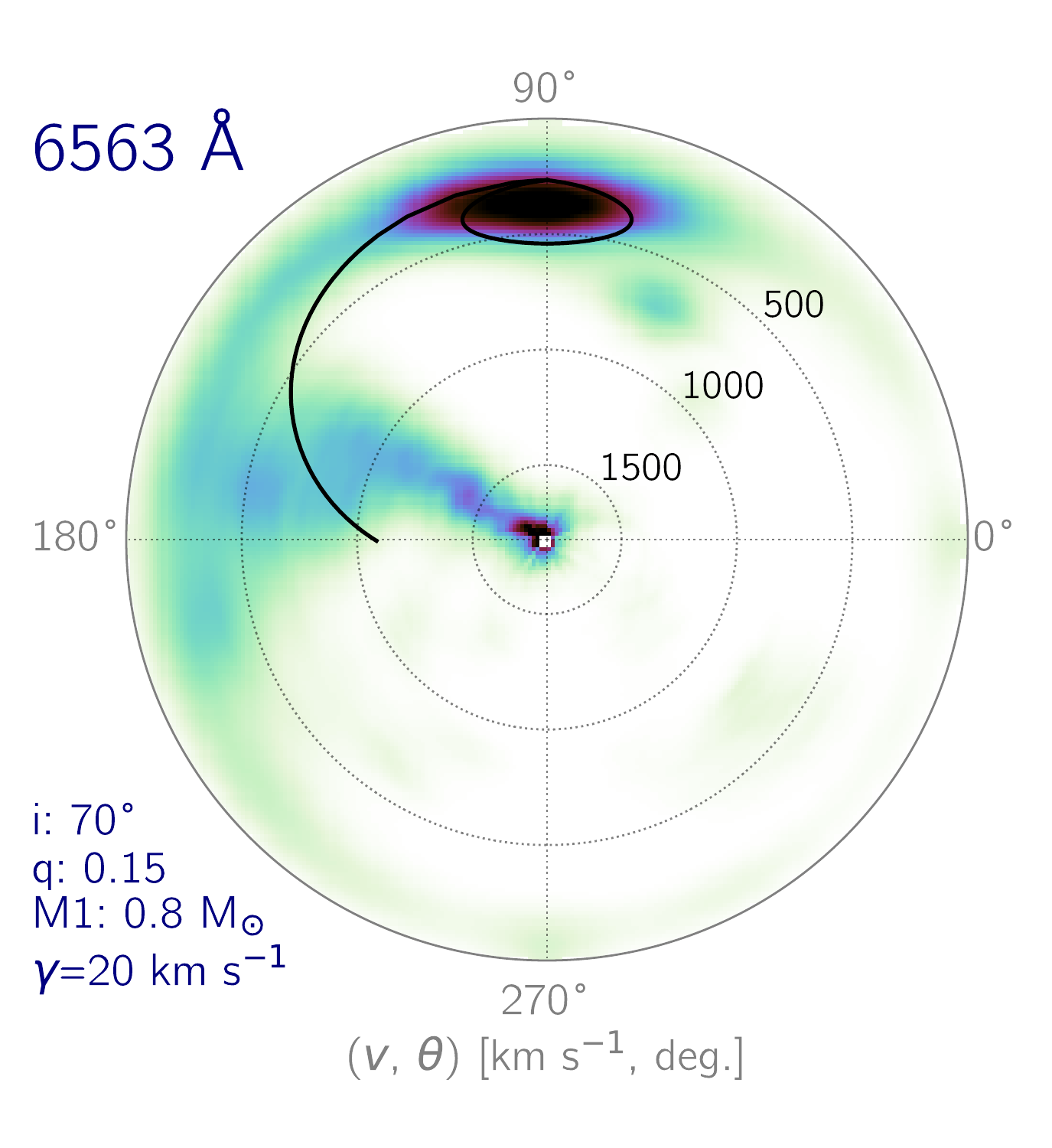}
        \includegraphics[width=\columnwidth]{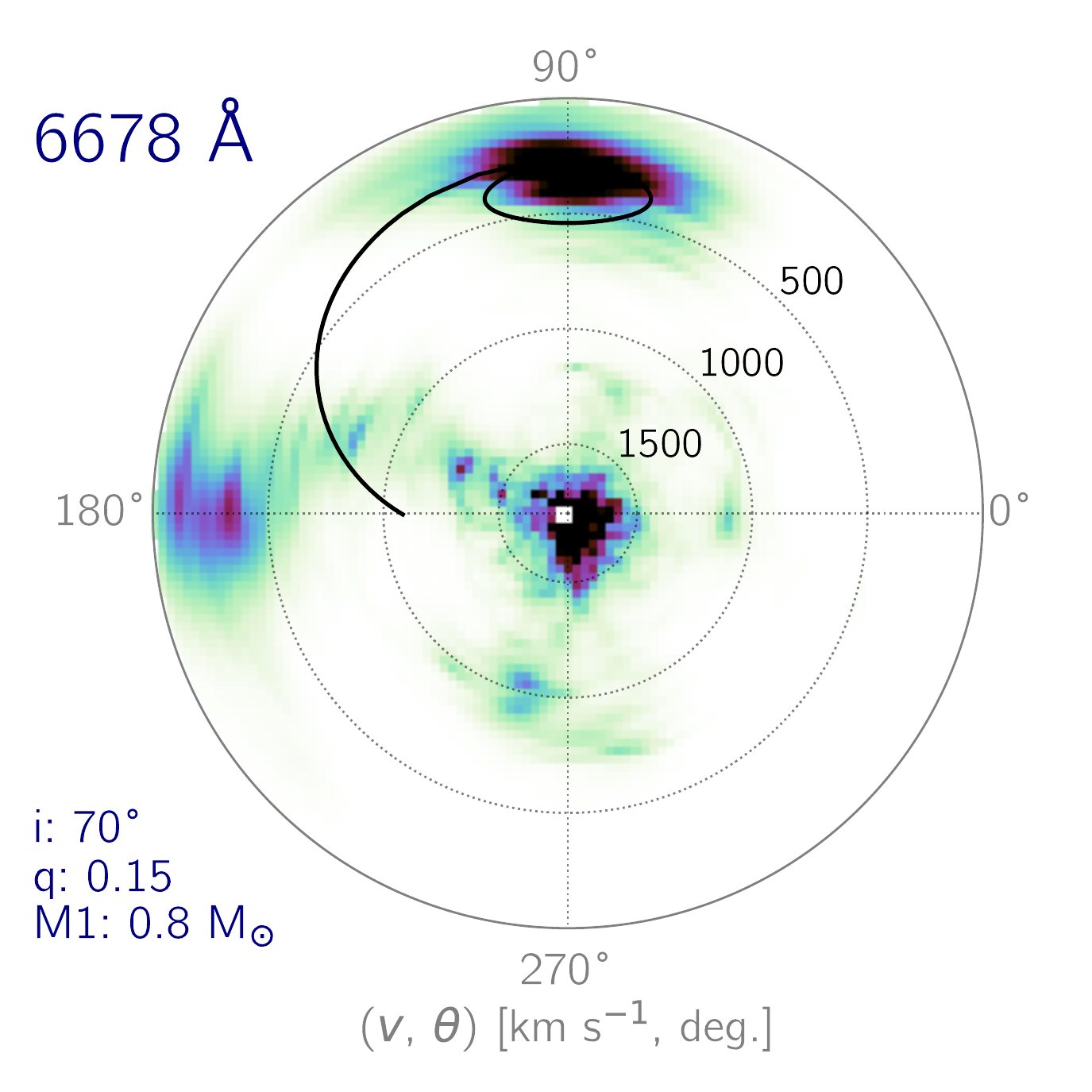}
        \caption{Doppler tomograms of the H$\alpha$ and He~I $\lambda$~6678~\AA\ lines during the 2019 LBT observations. For reference, we plot the theoretical trajectory of the ballistic stream for assumed values of the binary inclination $i$, the mass ratio $q$, and the WD mass. The accretion flow is best seen in the H$\alpha$ line, and its arc-like appearance suggests inflow from the secondary along a curved field line. We attribute the enhanced emission near $V=300$~km s$^{-1}$, 180$^{\circ}$ in the He~I tomogram to a threading region near the L1 point in which the magnetically confined flow preserves its initial orbital speed.}
        \label{fig:tomograms}
    \end{figure}

    To better study the accretion flow, we use inside-out Doppler tomography, as implemented by \citet{kotze}. Doppler maps of the H$\alpha$ and He~I $\lambda6678$\AA\ lines (Fig.~\ref{fig:tomograms}) strongly suggest that accretion occurred directly from the vicinity of the L1 point and that there was no ballistic accretion stream. Based on Fig.~1 in \citet{kotze}, which relates spatial coordinates to velocity coordinates, a ballistic stream would be expected to cause the curtain to appear somewhere between azimuths $180^{\circ}$and $270^{\circ}$, depending on the length of the ballistic stream as well as the azimuthal width of the accretion curtain. In Fig.~\ref{fig:tomograms}, however, the azimuth of the curtain is slightly less than $\sim180^{\circ}$, consistent with the expected azimuth for radial inflow from the vicinity of the L1 point.

    The He~I tomogram shows a discrete clump of enhanced emission at the orbital speed of the donor star's inner hemisphere but rotated by $+90^{\circ}$ compared to the position of the secondary. This is consistent with material near the secondary that is moving radially towards the WD instead of with the secondary. We interpret this as emission by material that has been threaded directly from the L1 point. If this supposition is correct, then it would appear that the threaded gas retains its orbital speed when it begins traveling along the field lines.

    \subsection{Spectral classification of the secondary}
    \label{sec:spectral_classification}
    
    The spectral type of the red dwarf is M9$\pm$1. We arrived at this classification by averaging the spectra obtained during the self-eclipse of the 2019 spectra, when the only continuum sources were the two stellar photospheres. We then modeled the self-eclipse spectrum by using a grid of WD and M-dwarf template spectra. To represent the WD, we used the \citet{koester} models\footnote{ The \citet{koester} models can be downloaded from \url{http://svo2.cab.inta-csic.es/theory/newov/index.php?model=koester2}.} and assumed that $\log{g} = 8.5$.  For the red dwarf, we relied upon the \citet{bochanski} template spectra of M and L0 dwarfs. \footnote{These templates are available at \url{https://github.com/jbochanski/SDSS-templates}.}
    Equipped with these models, we used $\chi^{2}$ minimization to fit them to the self-eclipse spectrum, allowing the model spectra to be scaled independently of each other. Our fitting procedure excluded parts of the spectrum that contained absorption from the WD or emission lines. The pairing that minimized $\chi^{2}$ was a WD with an effective temperature ($T_{eff}$) of 7250~K and an M9 dwarf, although several models with an M8 or L0 dwarf had only slightly higher $\chi^{2}$ values. We therefore estimate a donor-star spectral type of M9$\pm$1 and that $7000 \leq T_{eff} \leq 7750$ for the WD.

\begin{figure*}
        \centering
        \includegraphics[width=\textwidth]{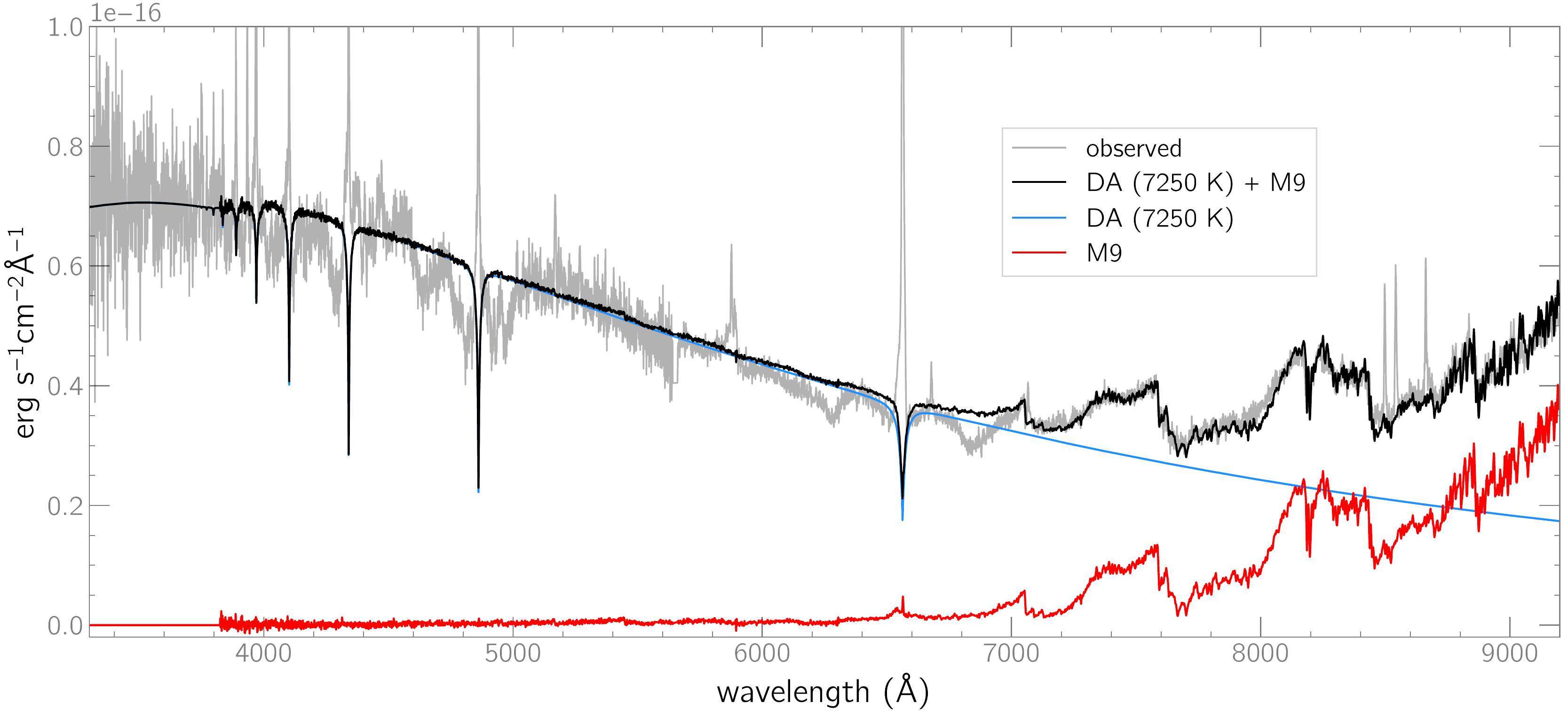}
        \caption{Decomposition of Tau~4's spectrum into the sum of two photospheric components. The observed spectrum is from the 2019 dataset and is the radial-velocity corrected mean of all spectra obtained during the self-eclipse, when there was no cyclotron radiation. Emission lines and WD absorption features were excluded from the fitting procedure, and the WD atmospheric model neglects magnetic effects. The combination that minimized $\chi^{2}$ (a M9 dwarf and a 7250~K WD) is shown, though an M8 or L0 dwarf can produce an almost identically good fit. We assumed a surface gravity of $\log(g) = 8.5$. }
        \label{fig:spectral_classification}
    \end{figure*}

    The decomposition of the spectrum in Fig.~\ref{fig:spectral_classification} allows us to crudely estimate the relative contributions of the two stellar photospheres in the \textit{Kepler} bandpass. By multiplying the two model spectra by the \textit{Kepler} transmission function and integrating, we find that the WD outshines the secondary by a factor of $\sim10$ in the \textit{Kepler} bandpass, although this estimate could be improved with a WD atmospheric model that includes magnetic effects.
    
    \subsection{Magnetic-field strengths from Zeeman emission and absorption components} \label{sec:zeeman}
    
    Both sets of spectra show obvious Zeeman absorption features from the photosphere of the WD, and the wavelengths of these features provides a surface-averaged field strength for the WD. \citet{zhao} computed the wavelengths of numerous H transitions across a wide range of magnetic-field strengths from 2.35~MG to 2350~MG, and we compared the wavelengths of the observed Zeeman features against their calculations. A field-strength of $15\pm2$~MG yields the best match between observed and predicted wavelengths. This is marginally higher than the field strength of 12.3~MG (no uncertainty specified) estimated by \citet{harrison} from cyclotron spectroscopy, but since these two techniques measure the field strength in different physical areas, it is not surprising that they are somewhat different.
    
    \begin{figure}
        \centering
        \includegraphics[width=\columnwidth]{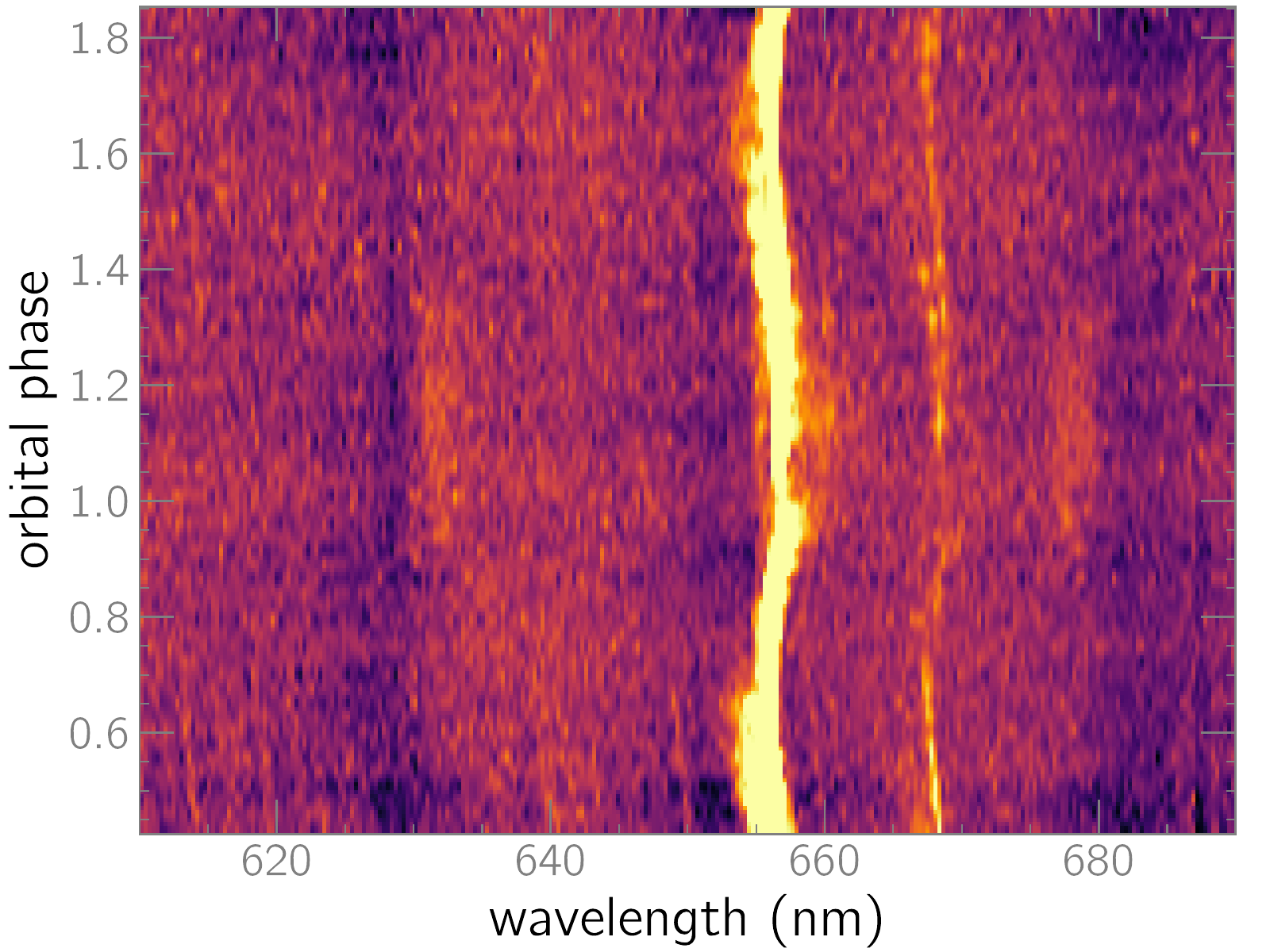}
        \includegraphics[width=\columnwidth]{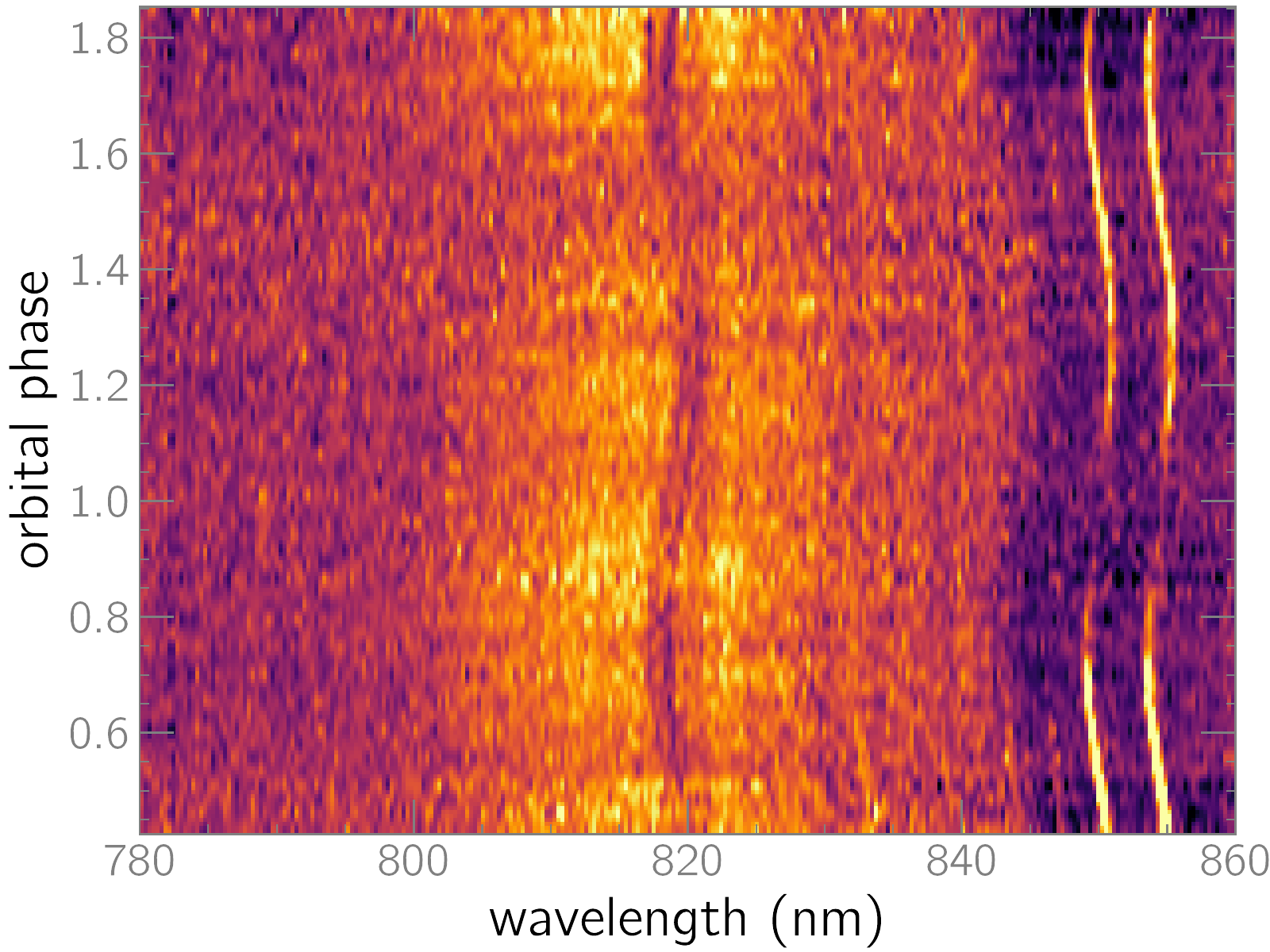}
        \caption{Trailed spectra from 2019. The continuum has been subtracted with a low-order polynomial. {\bf Top:} H$\alpha$ Zeeman emission lines. The emission is visible near 630~nm and 680~nm between orbital phases 0.9-1.3, corresponding with the orbital phases during which the accretion region is viewed most directly. The photospheric Zeeman absorption features of the WD, as well as the He~I $\lambda$667.8~nm emission line, are also visible. {\bf Bottom:} The near-infrared Na~I doublet at $\lambda\lambda$818.3, 819.5~nm, and nearby molecular absorption bands from TiO and VO. The two emission lines at right are Ca~II $\lambda\lambda$ 849.8, 854.2~nm. The contrast of the secondary's absorption features is lowest near orbital phase 0.5, which corresponds with the photometric minimum of the low-state orbital profile.}
        \label{fig:trailed_spectra}
    \end{figure}

    Remarkably, Tau~4 also shows Zeeman-split \textit{emission} lines that originate close to the WD's surface. The $\sigma^{-}$ and $\sigma^{+}$ components of the H$\alpha$ emission line appear as broad, low-contrast emission features that are visible between orbital phases 0.9-1.3 near 630~nm and 680~nm (Fig.~\ref{fig:trailed_spectra}), consistent with a field strength of $12\pm1$~MG.  We further identify the $\sigma^{-}$ emission component for H$\beta$, but the erratic continuum near H$\beta$ complicates the identification of the corresponding $\sigma^{+}$ feature. While Zeeman-split emission lines have been predicted for high-field (B~$> 100$~MG) polars \citep{fws00, ar_uma}, the expectation has been that they would be produced near the threading region, where the field strength is far lower \citep[\textit{e.g.}, 30~kG for AR UMa;][]{ar_uma}. However, in Tau~4, the Zeeman emission must originate very close to the WD; the measured field strength is very close to the WD's surface-averaged field strength, and the emission features disappear behind the limb of the WD for half of the orbital cycle. 
    
    This phenomenon bears some resemblance to Zeeman-split halo absorption, which has been observed to occur in other polars when the accretion region is viewed through a region of cool, infalling material very close to the WD \citep{halo, ferrario15}. Halo emission does not appear to have been reported previously in a polar, and while a heated halo could in principle produce Zeeman-split emission lines, the observed Zeeman emission in Tau~4 does not show the large-amplitude radial-velocity variations expected of infalling matter close to the WD (Fig.~\ref{fig:trailed_spectra}). To explain the apparent absence of these radial-velocity variations, we propose that the Zeeman emission features are the result of a localized temperature inversion in the atmosphere of the WD, caused by heating from particle bombardment. The calculations of \citet{wb92} show that the bombardment solution is expected to produce such an inversion in the accretion region, and if the Balmer-line-forming region were hotter than the underlying photosphere, the Balmer lines would be seen in emission. Because the Zeeman-emission region would be in the WD's atmosphere, limb-darkening effects might prevent the observation of the emitting region until it has moved closer to the meridian of the WD, explaining why the Zeeman emission is detectable only during the center of the cyclotron-bright part of the orbit.

\section{Discussion}

	\subsection{Accretion-state transitions in Tau 4}
	
	Although our dataset does not uniquely pinpoint the cause of Tau~4's accretion-state transitions, it does offer tantalizing clues that can be applied more broadly to the poorly understood nature of low states in polars.

\begin{figure}
	    \centering
	    \includegraphics[width=\columnwidth]{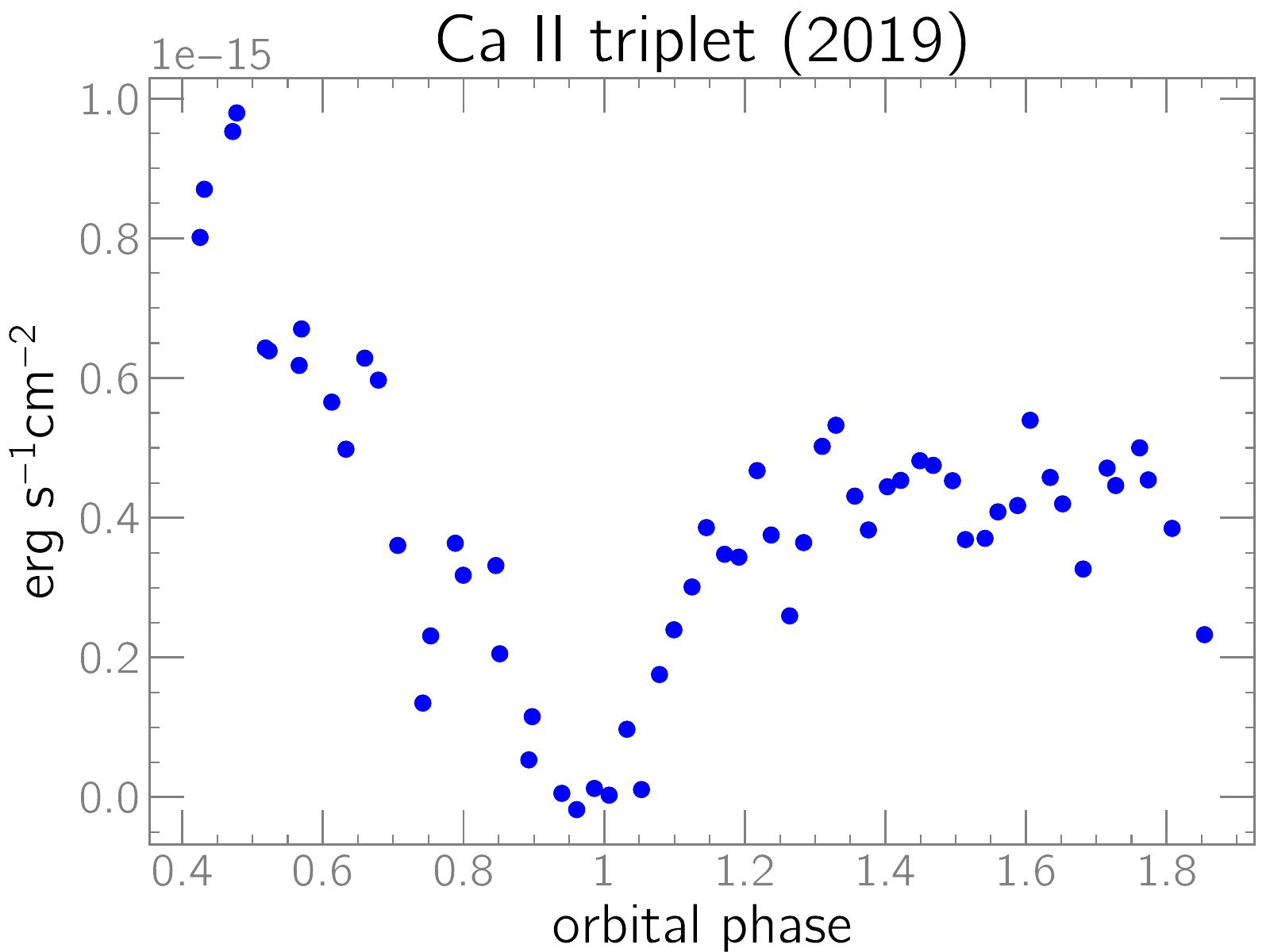}
	    \caption{Summed fluxes of the Ca~II near-infrared triplet, measured by fitting a Gaussian to each line. The line flux showed significant orbit-to-orbit variation without a stable maximum at superior conjunction, favoring stellar activity as the cause of the line emission as opposed to irradiation.}
	    \label{fig:Ca_line_flux}
	\end{figure}

	\citet{howell00} argued that in short-period CVs, the photosphere of the secondary significantly underfills the Roche lobe and that it is the chromosphere that fills out the Roche lobe. We do see enhanced chromospheric emission in the high state, particularly from the Na~I~D doublet at $\lambda\lambda5990, 5996$\AA\ and the Ca~II near-infrared triplet. In M dwarfs, the Na~I~D doublet probes conditions in the chromosphere. \citet{andretta} and \citet{sd98} found that it is sensitive to the thickness of the chromosphere, although our spectra lack both the resolution and the signal-to-noise ratio to apply the modeling used in those studies. In contrast, there was no detectable Na~D emission during the low state. While irradiation by the WD could produce emission lines on the inner hemisphere, there are several lines of circumstantial evidence against this scenario. If the line-emission were induced by irradiation from the WD, we would expect the irradiation to be present during the low state, too. Moreover, the estimated temperature of the WD in Tau~4 is less than 8,000~K (Sec.~\ref{sec:spectral_classification}), and \citet{howell_book} showed that even for a 10,000~K WD, there would be insufficient UV irradiation to have a significant effect on the secondary. Additionally, while an accretion shock during the high state would irradiate the secondary with X-rays, the defining characteristic of the bombardment state is the very absence of such a shock. Finally, as \citet{mason} pointed out, we would expect irradiation to cause the chromospheric line flux to be largest near superior conjunction, when the inner hemisphere of the secondary is best viewed, and to be repeatable from orbit-to-orbit. However, Tau~4's Ca~II lines contradict these predictions (Fig.~\ref{fig:Ca_line_flux}). Collectively, these arguments disfavor irradiation of the secondary as the cause of the chromospheric emission.
	
	Additionally, \citet{mason} modelled low-state spectra of the polar VV~Pup to determine whether stellar activity or irradiation was the more likely culprit for the observed low-state emission spectrum. VV~Pup's orbital period, inclination, and secondary spectral type are similar to Tau~4, making it a reasonable point of comparison. While they found that irradiation by a 11,900~K WD could produce emission by a series of low-excitation metal lines near 500~nm, emission from Na, He, and the Balmer lines only became apparent at much higher WD temperatures (20,000~K). While they noted that a localized hotspot on a much cooler WD could produce H emission, the predicted Balmer decrement was inverted, which is contrary to our observations of Tau~4. While \citet{mason} did caution that their simulated spectra were dependent on several poorly constrained factors, their modeling provides circumstantial evidence against irradiation-induced emission in Tau~4.
	
	The chromospheric model of \citet{howell00} can qualitatively explain (1) why Tau~4 was in the bombardment regime and (2) why threading occured very close to the L1 point: both are consequences of the chromosphere's very low density. These behaviors are possible with a very low \mdot, although detailed theoretical modeling would make it possible to explore this scenario more rigorously.

	\subsubsection{Comparison to \lsqfull}

	The observed timescale of Tau~4's accretion-state transition is  comparable to that observed by \citet{fuchs} in ground-based photometry of two low-to-high-state transitions in \lsqfull, a polar with a similar orbital period, magnetic-field strength, and donor-star spectral type. The better-sampled of the two transitions in \lsq\ occurred over the course of 3~d ($\sim$45 binary orbits), while the second took $< 5$~d. At face value, both estimates are somewhat faster than Tau~4's $\sim6$~d transition. However, our timescale for Tau~4's transition is based on when the very first accretion events took place, whereas the ground-based light curves in \citet{fuchs} were unable to pinpoint when these occurred for \lsq. With this caveat in mind, we conclude that the transitions in the two systems took the same amount of time.  
	
	It is worth noting that \lsq\ appears to be a bombardment-regime accretor based on the spectra in \citet{fuchs}. Similar to Tau~4, it showed exceptionally weak He~II $\lambda4686$\AA\ emission, and the maximum redshift of its high-velocity emission was observed shortly after inferior conjunction. Although \citet{fuchs} proposed that this emission from from a ballistic accretion stream, we instead propose that it might have been from a magnetically confined accretion flow from the L1 point (again similar to Tau~4). If so, our earlier argument about chromospheric Roche-lobe overflow and its predisposition towards both the bombardment regime and L1 threading would apply equally to \lsq.

    \subsection{The low-state orbital light curve}
    
    The low-state orbital light curve of Tau~4 is nearly identical to that of \lsq. Both consist of a double-humped modulation with unequal minima and maxima. In both systems, the primary minimum occurs at orbital phase 0.5, exactly a quarter-orbit after the primary maximum and a quarter-orbit before the secondary maximum. As \citet{fuchs} pointed out, the phasing is strongly suggestive of ellipsoidal variability, but after correcting for dilution by the WD, the ellipsoidal variations in \lsq\ would have a fractional amplitude of 50\% of the system's total optical flux---far above the theoretical maximum predicted for ellipsoidal variations (see their Appendix A1). Given the meager contribution of Tau~4's secondary in the \textit{Kepler} bandpass (Sec.~\ref{sec:spectral_classification}), the same theoretical arguments invoked by \citet{fuchs} would disfavor ellipsoidal variations as the source of Tau~4's low-state orbital variability.

    Instead, \citet{fuchs} proposed that hotspots on the WD were responsible for the observed low-state variability in \lsq. While this explanation is plausible for one system, it would be an unlikely coincidence that WD hotspots could produce such identically phased low-state variability in different systems. Hotspots induced by previous accretion should appear at different orbital phases in different polars, because the longitude of the WD's magnetic axis will differ from system-to-system.
    Tau~4 and \lsq\ are very different in that respect; in the former, the accretion region crosses the meridian at orbital phase $\sim0.15$, while this occurs at orbital phase $\sim0.75$ in the latter. Because the low-state profiles of the two systems are nevertheless identical, we believe that hotspots are an unlikely explanation.
    
    For similar reasons, it is unlikely that low-state cyclotron radiation, similar to that observed in AR~UMa \citep{howell_ar_uma}, could produce identical orbital light curves in both Tau~4 and \lsq. As with the hotspot scenario, the photometric maxima would be expected at different orbital phases in different systems. Indeed, in the case of AR~UMa, the phasing of the low-state modulation (Fig.~4 in \citealt{howell_ar_uma}) is dissimilar to that observed in both Tau~4 and \lsq, which suggests that a different mechanism produces the low-state orbital variability in AR~UMa. 
    
    Instead, we believe that ellipsoidal variations are responsible for the low-state orbital light curve in Tau~4, and our spectra show why this might be. In the bottom panel of Fig.~\ref{fig:trailed_spectra}, we present a trailed spectrum in the vicinity of the near-infrared Na~I doublet. The secondary's absorption features (Na~I as well as nearby molecular bands) almost fade into the continuum near orbital phase 0.5---the same phase as the primary minimum in the low state. This suggests that the secondary causes the observed photometric variability, because if the minimum were instead caused by the disappearance of a hotspot, the reduced dilution of the secondary's contribution would boost the contrast of its spectral features---the exact opposite of what we observe.
    
    Even more fundamentally, the phasing of ellipsoidal variations should be identical in different systems, which would explain the excellent phase agreement between the low-state orbital profiles of both Tau~4 and \lsq. Although the amplitude of the observed variability is too large according to the theoretical work cited by \citet{fuchs}, we propose that the underlying models might need to be refined to better describe the peculiarities of short-period CVs. For example, it could be that detailed atmospheric models of the secondary are necessary to accurately describe its ellipsoidal variations.

	\subsection{On the absence of low-state flares}

	Low-state polars offer an excellent opportunity to examine the activity of the secondary star, as a flare would result in an accretion spurt onto the WD on approximately the free-fall timescale. Previous studies of low-state polars have suggested that such events might be common. \citet{pandel} reported X-ray observations of two low-state polars, VV~Pup and V393~Pav, and in parts of four binary orbits, they detected numerous X-ray flares consistent with temporary spurts of accretion. Similarly, UZ~For showed three UV flares across parts of 5 consecutive orbits during a low-state XMM-Newton observation \citep{pandel02}; simultaneous X-ray observations were available for one of the flares and strongly suggested that it was a transient accretion event \citep{still01, pandel02}. Another observation of low-state flaring comes from \citet{bb00}, who identified 5 flares in just under 8 hours of observations of AM Her.
	
	\begin{figure}
        \centering
        \includegraphics[width=\columnwidth]{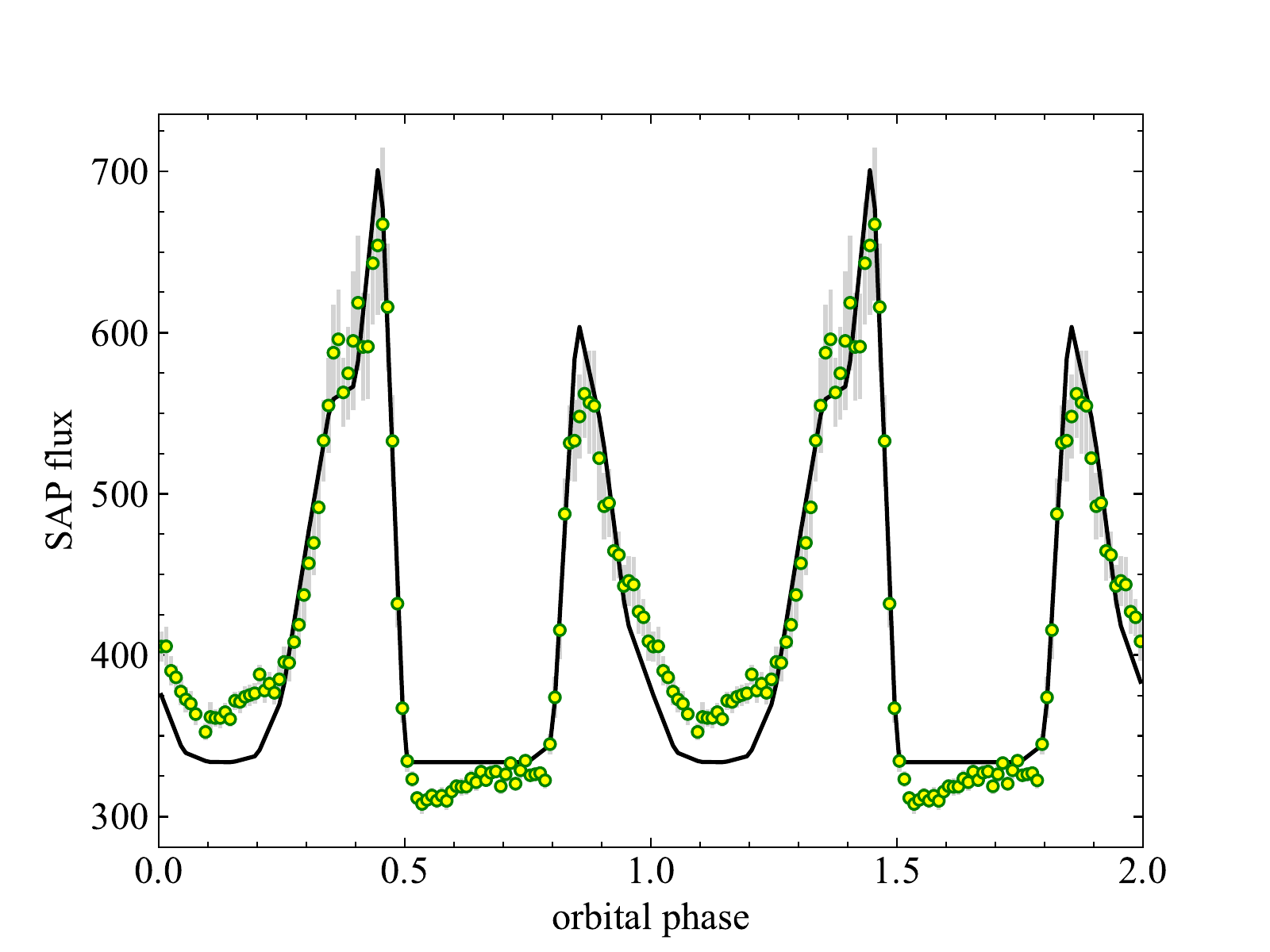}
        \caption{Phase-averaged orbital profile of Tau~4 during the high state (points) and the best-fit cyclotron model from CYCLOPS (solid line). The model parameters are described in Sec.~\ref{sec:cyclops}.  \label{fig:cyclops_model}}
    \end{figure}

    Against this backdrop, it is remarkable that Tau~4 showed just a single flare during the $>1000$ orbits observed during the low state. As \citet{pandel} pointed out, these flares almost certainly result from eruptive activity that causes a short-lived spurt of mass transfer from the secondary. It is possible that the much higher rate of flaring in VV Pup and V393 Pav is the result of these systems' low states not being representative of all low state in polars. For example, low-state spectroscopy of VV~Pup by \citet{howell06} showed unmistakable cyclotron emission, suggesting a low-state mass-transfer rate of $10^{-13}$M$_{\odot}$ yr$^{-1}$. In contrast, we do not see any evidence of accretion in Tau~4 during its low state, and the absence of flaring suggests that there are multiple types of low states in AM~Herculis systems.

    \subsection{Cyclotron modeling} \label{sec:cyclops}
	
	We analyzed the cyclotron light curve of Tau~4 during the high state using CYCLOPS \citep{costa, silva}, which models the post-shock region of magnetic CVs. CYCLOPS has nine parameters that define the geometrical and physical properties of the post-shock region. The current version solves the stationary one-dimensional hydrothermodynamic equations, allowing gas to cool via cyclotron and bremsstrahlung radiative processes \citep{O19}. It presupposes the existence of a shock, and to our knowledge, it has not been tested previously with a bombardment-regime polar like Tau~4.
	
	We searched for the best-fit model by minimizing $\chi^{2}$ at three model wavelengths (477~nm, 591~nm, 754~nm) selected to approximate the \textit{Kepler} bandpass. Initially, we kept all parameters free to define the accretion geometry. Then, with the geometrical parameters fixed, we looked for the physical parameters with more accuracy.
 
    In the absence of polarimetry, we could not arrive at a unique CYCLOPS solution. A number of models yielded comparably low values of $\chi^{2}$, displaying a degeneracy between the white dwarf mass (M) and $\dot{M}$; the larger values of M are obtained for the smaller values of $\dot{M}$, and vice-versa. If we assume that the WD has a typical mass for a CV, then the solutions with heavier WDs are more likely. As an example of one of these non-unique solutions, we show in Fig.~\ref{fig:cyclops_model} the model with M~=0.61M$_{\odot}$ and $\dot{M}$ =10$^{-12}$~M$_{\odot}$~yr$^{-1}$~=$\sim$10$^{11}$~kg~s$^{-1}$. The inclination is 67$^{\circ}$, and the magnetic-field axis makes an angle of 63$^{\circ}$ with the rotation axis. In the emitting region, the magnetic field ranges from 15.4~MG to 17.5~MG. It is encouraging that the magnetic-field strength is in reasonable agreement with the surface-averaged field strength of $15\pm2$~MG found from Zeeman splitting (see Sec.~\ref{sec:zeeman}). Moreover, the model's $\dot{M}$ is over an order of magnitude smaller than the rate expected from the optimal CV evolutionary track from \citet{knigge}, which predicts $\dot{M} \sim 4\times10^{12}$~kg~s$^{-1}$ at this orbital period for a Roche-lobe-filling secondary. This is consistent with our argument from Sec.~\ref{sec:spectra2019} that $\dot{M}$ must have been very low during the high state observed by \textit{Kepler}.

\section{Conclusion}

	The \textit{K2} light curve of Tau~4 captured a serendipitous transition from a low-accretion state into a high state. The transition occured gradually over the course of approximately 90 orbits, and based on subsequent spectroscopy of Tau~4, we argue that the secondary's chromosphere is the likely source of the matter that was transferred. We find evidence of enhanced activity in the secondary during the high state, which supports proposals that in short-period polars, the chromosphere fills the Roche lobe and that low states are episodes of diminished chromospheric activity. Furthermore, our spectra show that Tau~4 was accreting in the bombardment regime and that threading occured in the vicinity of the L1 point. Finally, we report the first detection of Zeeman-split emission lines in a polar and attribute them to a localized temperature inversion in the WD's atmosphere, possibly caused by heating from bombardment-regime accretion.

\acknowledgements

We thank Lilia Ferrario for a helpful discussion about Zeeman splitting.

M.R.K. acknowledges support from the ERC under the European Union's Horizon 2020 research and innovation programme (grant agreement No. 715051; Spiders).

\software{Astropy \citep{astropy13, astropy18}, doptomog \citep{kotze}, CYCLOPS \citep{costa, silva} lightkurve \citep{lightkurve} }


\begin{thebibliography}{0}
	
		
	\bibitem[Achilleos et al.(1992)]{halo} Achilleos, N., Wickramasinghe, D.~T., \& Wu, K.\ 1992, \mnras, 256, 80
	
	\bibitem[Andretta et al.(1997)]{andretta} Andretta, V., Doyle, J.~G., \& Byrne, P.~B.\ 1997, \aap, 322, 266	
		
	\bibitem[Astropy Collaboration et al.(2013)]{astropy13} Astropy Collaboration, Robitaille, T.~P., Tollerud, E.~J., et al.\ 2013, \aap, 558, A33
	
	\bibitem[Astropy Collaboration et al.(2018)]{astropy18} Astropy Collaboration, Price-Whelan, A.~M., Sip{\'{o}}cz, B.~M., et al.\ 2018, \aj, 156, 123
	
	\bibitem[Bailer-Jones et al.(2018)]{bj18} Bailer-Jones, C.~A.~L., Rybizki, J., Fouesneau, M., et al.\ 2018, \aj, 156, 58.
	
	\bibitem[Bochanski et al.(2007)]{bochanski} Bochanski, J.~J., West, A.~A., Hawley, S.~L., et al.\ 2007, \aj, 133, 531
	
	\bibitem[Bonnet-Bidaud et al.(2000)]{bb00} Bonnet-Bidaud, J.~M., Mouchet, M., Shakhovskoy, N.~M., et al.\ 2000, \aap, 354, 1003
	
	\bibitem[Costa \& Rodrigues(2009)]{costa} Costa, J.~E.~R., \& Rodrigues, C.~V.\ 2009, \mnras, 398, 240
	
	\bibitem[Chambers et al.(2016)]{chambers} Chambers, K.~C., Magnier, E.~A., Metcalfe, N., et al.\ 2016, arXiv e-prints, arXiv:1612.05560
	
	\bibitem[Cropper(1990)]{cropper} Cropper, M.\ 1990, \ssr, 54, 195
	
	\bibitem[Drake et al.(2009)]{drake} Drake, A.~J., Djorgovski, S.~G., Mahabal, A., et al.\ 2009, \apj, 696, 870
	
	\bibitem[Ferrario et al.(2001)]{fws00} Ferrario, L., Wickramasinghe, D.~T., \& Schmidt, G.~D.\ 2001, Magnetic Fields Across the Hertzsprung-russell Diagram, 463

	
    \bibitem[Ferrario et al.(2015)]{ferrario15} Ferrario, L., de Martino, D., \& G{\"a}nsicke, B.~T.\ 2015, \ssr, 191, 111
	
	\bibitem[Fuchs et al.(2016)]{fuchs} Fuchs, J.~T., Dunlap, B.~H., Dennihy, E., et al.\ 2016, \mnras, 462, 2382.
	
	\bibitem[Gaia Collaboration et al.(2016)]{gaia} Gaia Collaboration, Prusti, T., de Bruijne, J.~H.~J., et al.\ 2016, \aap, 595, A1
	
	\bibitem[Gaia Collaboration et al.(2018)]{dr2} Gaia Collaboration, Brown, A.~G.~A., Vallenari, A., et al.\ 2018, ArXiv e-prints, arXiv:1804.09365
	
	\bibitem[G\"ansicke et al.(2001)]{gansicke} G\"ansicke, B., Fischer, A., Silvotti, R., \& de Martino, D. 2001, \aap, 372, 557
	
	\bibitem[Hakala et al.(2019)]{hakala} Hakala, P., Ramsay, G., Potter, S.~B., et al.\ 2019, \mnras, 486, 2549
	
	\bibitem[Harrison(2018)]{harrison} Harrison, T.~E.\ 2018, RNAAS, 2, 5
	
	\bibitem[Hill et al.(2019)]{hill} Hill, K., Littlefield, C., Garnavich, P., \& Szkody, P.\ 2019, RNAAS, 3, 103
	
	\bibitem[Howell et al.(2000)]{howell00} Howell, S.~B., Ciardi, D.~R., Dhillon, V.~S., et al.\ 2000, \apj, 530, 904
	
	\bibitem[Howell et al.(2001)]{howell_ar_uma} Howell, S.~B., Gelino, D.~M., \& Harrison, T.~E.\ 2001, \aj, 121, 482
	
	\bibitem[Howell et al.(2006)]{howell06} Howell, S.~B., Harrison, T.~E., Campbell, R.~K., et al.\ 2006, \aj, 131, 2216
	
	\bibitem[Howell et al.(2008)]{howell} Howell, S.~B., Harrison, T.~E., Huber, M.~E., et al.\ 2008, \aj, 136, 2541
	
	\bibitem[Howell(2008)]{howell_book} Howell, S.~B.\ 2008, Astrophysics and Space Science Library, 147
	
	\bibitem[King \& Cannizzo(1998)]{kc98} King, A.~R., \& Cannizzo, J.~K.\ 1998, \apj, 499, 348
	
	\bibitem[Kochanek et al.(2017)]{kochanek} Kochanek, C.~S., Shappee, B.~J., Stanek, K.~Z., et al.\ 2017, PASP, 129, 104502
	
	\bibitem[Koester(2010)]{koester} Koester, D.\ 2010, \memsai, 81, 921
	
	\bibitem[Kotze et al.(2016)]{kotze} Kotze, E.~J., Potter, S.~B., \& McBride, V.~A.\ 2016, \aap, 595, A47
	
	\bibitem[Knigge et al.(2011)]{knigge} Knigge, C., Baraffe, I., \& Patterson, J.\ 2011, \apjs, 194, 28
	
	\bibitem[Kuijpers \& Pringle(1982)]{kp82} Kuijpers, J., \& Pringle, J.~E.\ 1982, \aap, 114, L4
	
	\bibitem[Littlefair et al.(2005)]{littlefair} Littlefair, S.~P., Dhillon, V.~S., \& Mart{\'\i}n, E.~L.\ 2005, \aap, 437, 637
	
	\bibitem[Littlefield et al.(2018)]{j1321} Littlefield, C., Garnavich, P., Hoyt, T.~J., et al.\ 2018, \aj, 155, 18	
	
	\bibitem[Littlefield et al.(2019)]{littlefield} Littlefield, C., Garnavich, P., Mukai, K., et al.\ 2019, \apj, 881, 141
	
	\bibitem[Livio \& Pringle(1994)]{lp94} Livio, M., \& Pringle, J.~E.\ 1994, \apj, 427, 956
	
	\bibitem[Mason et al.(2008)]{mason} Mason, E., Howell, S.~B., Barman, T., et al.\ 2008, \aap, 490, 279
	
	
	\bibitem[Motch et al.(1996)]{motch} Motch, C., Haberl, F., Guillout, P., et al.\ 1996, \aap, 307, 459
	
	\bibitem[Oliveira et al.(2019)]{O19} Oliveira A.~S., Rodrigues C.~V., Palhares M.~S., Diaz M.~P., Belloni D., Silva K.~M.~G.\ 2019, \mnras, 489, 4032
	
	\bibitem[Pandel \& C{\'o}rdova(2002)]{pandel02} Pandel, D., \& C{\'o}rdova, F.~A.\ 2002, \mnras, 336, 1049
	
	
	\bibitem[Pandel \& C\'ordova(2005)]{pandel} Pandel, D., \& C\'ordova, F. 2005, \apj, 620, 416.
	
	\bibitem[Pogge et al.(2010)]{pogge} Pogge, R.~W., Atwood, B., Brewer, D.~F., et al.\ 2010, \procspie, 77350A
	
	\bibitem[Schmidt et al.(2001)]{ar_uma} Schmidt, G.~D., Ferrario, L., Wickramasinghe, D.~T., et al.\ 2001, \apj, 553, 823
	
	\bibitem[Schmidt et al.(2005)]{schmidt} Schmidt, G.~D., Szkody, P., Vanlandingham, K.~M., et al.\ 2005, The Astrophysical Journal, 630, 1037.
	
	
	\bibitem[Silva et al.(2013)]{silva} Silva, K.~M.~G., Rodrigues, C.~V., Costa, J.~E.~R., et al.\ 2013, \mnras, 432, 1587
	
	\bibitem[Shappee et al.(2014)]{shappee} Shappee, B.~J., Prieto, J.~L., Grupe, D., et al.\ 2014, \apj, 788, 48.
	
	
	\bibitem[Short \& Doyle(1998)]{sd98} Short, C.~I., \& Doyle, J.~G.\ 1998, \aap, 336, 613
	
	
	
    \bibitem[Still \& Mukai(2001)]{still01} Still, M., \& Mukai, K.\ 2001, \apjl, 562, L71
    
    \bibitem[Szkody et al.(2008)]{s08} Szkody, P., Linnell, A.~P., Campbell, R.~K., et al.\ 2008, \apj, 683, 967
	
	\bibitem[Vin\'icius et al.(2018)]{lightkurve} Vin\'icius, Z., Barentsen, G., Hedges, C., Gully-Santiago, M., \& Cody, A.~M. 2018, lightkurve, v. 1.0b16, Zenodo, doi:10.5281/zenodo.1181928
	
	\bibitem[Woelk \& Beuermann(1992)]{wb92} Woelk, U., \& Beuermann, K.\ 1992, \aap, 256, 498

	
	\bibitem[Zhao \& Liu(2019)]{zhao} Zhao, L.~B., \& Liu, F.~L.\ 2019, \mnras, 486, 3849


	
	
\end{thebibliography}
\end{document}